\pdfoutput=1
\documentclass[aps, prl,
superscriptaddress,
amsmath,amssymb,
longbibliography,
reprint
]{revtex4-2}
\usepackage{graphicx}
\usepackage{textcomp, gensymb}
\usepackage{indentfirst}

\usepackage[normalem]{ulem} %For sout command

\usepackage{bm}
\usepackage{dcolumn}
\usepackage{amsmath}
\usepackage{amssymb}
\usepackage[utf8]{inputenc}
\usepackage[T1]{fontenc}
\usepackage{etoolbox}
\usepackage{hyperref}
\usepackage{mhchem}
\hypersetup{
    colorlinks=true,
    allcolors=blue
    }
\usepackage{xr}

\begin{document}

%%%%%%%%%%%%%%%%%%%%%%%%%%%%%%%%%%%%%%%%%%%%%%%%%%%%%%%%%%%%%%%%%%%%%%%%%%%%%%%
%\title{Optical tuning of diamond color centers charge state via photocatalytic surface oxidation}

\title{Photocatalytic Control of Diamond Color Center Charge States via Surface Oxidation}

%Laser-Induced Surface Oxidation for Charge State Control of Diamond Color Centers

\author{Minghao Li}
\email{minghao.li@unibas.ch}
\thanks{These authors contributed equally.}
\affiliation{Department of Physics, University of Basel, CH-4056 Basel, Switzerland}
\author{Josh A. Zuber}
\thanks{These authors contributed equally.} 
\affiliation{Department of Physics, University of Basel, CH-4056 Basel, Switzerland}
\affiliation{Swiss Nanoscience Institute, University of Basel, CH-4056 Basel, Switzerland}
\author{Marina Obramenko}
\affiliation{Department of Physics, University of Basel, CH-4056 Basel, Switzerland}
\affiliation{Swiss Nanoscience Institute, University of Basel, CH-4056 Basel, Switzerland}
\author{Patrik Tognina}
\affiliation{Department of Physics, University of Basel, CH-4056 Basel, Switzerland}
\author{Andrea Corazza}
\affiliation{Department of Physics, University of Basel, CH-4056 Basel, Switzerland}
\author{Marietta Batzer}
\affiliation{Department of Physics, University of Basel, CH-4056 Basel, Switzerland}
\author{Marcel.li Grimau Puigibert}
\affiliation{Department of Physics, University of Basel, CH-4056 Basel, Switzerland}
\author{Jodok Happacher}
\affiliation{Department of Physics, University of Basel, CH-4056 Basel, Switzerland}
\author{Patrick Maletinsky}
\email{patrick.maletinsky@unibas.ch}
\affiliation{Department of Physics, University of Basel, CH-4056 Basel, Switzerland}
\affiliation{Swiss Nanoscience Institute, University of Basel, CH-4056 Basel, Switzerland}

\date{Jun. 10, 2025}

%%%%%%%%%%%%%%%%%%%%%%%%%%%%%%%%%%%%%%%%%%%%%%%%%%%%%%%%%%%%%%%%%%%%%%%%%%%%%%%

%\begin{abstract}
%Surface functionalization plays a critical role in enabling precise charge state control of near-surface color centers in diamond, which is essential for many quantum technologies. 
%However, conventional chemical terminations often lack the tunability needed to stabilize the desired charge state.
%Here, we present a deterministic and nonvolatile method for continuously modulating surface termination via laser-induced oxidation of H-terminated surface of diamond nanopillars. 
%By monitoring SiV$^-$ photoluminescence as a charge state indicator, we reveal the microscopic mechanism of this photocatalytic process by a thorough analysis of its photon flux dependence. 
%Further investigation of the photon energy dependence of the reaction rate identifies charge-cycling of native defects, such as NV centers and divacancies, as the source of photogenerated holes driving photocatalysis. 
%While our study focuses on SiV centers, the method is broadly applicable to other color centers and host materials, offering a novel approach for on-demand charge state control and surface engineering in solid state quantum devices.
%\end{abstract}

\begin{abstract}
Color center spins in diamond nanostructures are a key resource for emerging quantum technologies. 
Their innate surface proximity makes precise control of diamond surface chemistry essential for optimizing their functionality and charge states. 
However, conventional surface functionalization methods typically lack the tunability and efficiency required for robust charge-state control.
Here, we introduce a deterministic, nonvolatile technique for continuously and efficiently tuning diamond's surface termination via laser-induced oxidation of H-terminated diamond nanopillars.
By tracking SiV$^-$ photoluminescence as a charge-state proxy, we uncover the microscopic mechanism of this photocatalytic process through a systematic photon-flux and -energy analysis, where we identify charge-cycling of native defects as sources of optically generated holes driving the desired surface oxidation.
Our results suggest that our method applies broadly to other color centers and host materials, offering a versatile tool for on-demand charge-state control and surface engineering in solid-state quantum devices.
\end{abstract}

\maketitle

\section{Introduction}

Fluorescent solid-state spin defects in wide-band-gap semiconductors have emerged as a promising platform for quantum sensing, communication, and computing due to their unique optical and spin properties\,\cite{wolfowicz2021,gao2015,atature2018}. 
To ensure optimal performance in such applications, it is essential that the charge state of the defect is well defined and stable. 
For bulk defects, charge-state stabilization can be achieved directly via Fermi-level engineering by doping the host crystal\,\cite{collins2002,rose_observation_2018}. However, many quantum applications, such as sensing, require defects to be incorporated into nanostructures or to reside near the diamond surface. 
For such shallow spin defects, the charge state can additionally be tuned by surface Fermi level pinning, which can be directly manipulated by chemical surface functionalizations\,\cite{strobel2004, hauf_chemical_2011, zhang_neutral_2023,rodgers_diamond_2024}.
Thus, achieving a specific surface termination is essential for stabilizing the desired charge state of shallow defects to guarantee performance and reliability in a range of applications. 

Among the various known solid-state spin defects, neutral and negatively charged silicon vacancies (SiV$^0$ and SiV$^-$) in diamond stand out due to their excellent optical and spin properties\,\cite{rogers2014, sipahigil2014, rose_observation_2018}.
The SiV$^-$ color center provides narrow, bright optical transitions\,\cite{zuber2023shallow} but exhibits rapid spin decoherence at temperatures exceeding a few Kelvin, whereas SiV$^0$ maintains long spin coherence times at elevated temperatures ($\sim 10\,$K) while retaining coherent optical emission characteristics\,\cite{rose_observation_2018}. 
For near-surface SiVs, it was recently shown that the charge state can be switched between neutral and negative by applying hydrogen and oxygen surface terminations, respectively\,\cite{zhang_neutral_2023}. 
However, preliminary observations also indicate that SiV$^0$ is not well stabilized under a completely H-terminated surface, and that such SiV's tend to transition to SiV$^+$, as suggested by significant blinking and spectral wandering\,\cite{CorazzaPrivate_2025}.
This underscores a more fundamental knowledge gap, namely that surface termination with a single chemical species may fail to reliably produce the desired charge state.
Indeed, theoretical studies have predicted that a mixed-species surface termination might yield the most ideal stabilizing conditions for certain color center spins close to the surface\,\cite{kaviani_proper_2014}.
Deterministic strategies to obtain such optimal surface terminations, however, have yet to be discovered and would constitute a key ingredient for further advancing the application potential of color center spins, especially in nanostructured systems where surface effects dominate.

Here, we demonstrate nonvolatile, optical tuning of the SiV charge state in diamond nanopillars, suggesting an efficient and deterministic method to achieve such termination.
Our work leverages the reactivity of H-terminated diamond surfaces under visible light to induce laser-driven surface oxidation. 
While prior studies\,\cite{pederson_optical_2024, rodgers_diamond_2024} have demonstrated the proof-of-concept of this approach, the microscopic mechanism underlying these laser-induced processes remains underexplored and the efficiency and timescales involved have been a severely limiting factor thus far. 
We address these challenges by first carefully modeling the charge state time evolution and by analyzing the photon flux dependence of the associated rate constant, confirming the photocatalytic nature of our laser-driven surface oxidation process.
Then, through a systematic study of the wavelength dependence of the photocatalytic reaction rates, we reveal charge-cycling, native diamond defects as the main source of photo-generated holes driving the reaction.
Furthermore, we show that diamond nanostructures can enhance the photocatalytic $\ce{C-H}$ bond activation reaction rates by orders of magnitude over previous methods\,\cite{rodgers_diamond_2024}.
While our study is conducted on the exemplary case of the SiV center in diamond, our results indicate the general applicability of this approach to other color centers and possibly other host materials, thereby offering a robust and versatile route for deterministic charge state control for a range of quantum technology applications.

\section{Results}
%\subsection{SiV Charge state conversion under laser illumination}

 \begin{figure}[t]
    \centering
    \includegraphics[width=85mm]{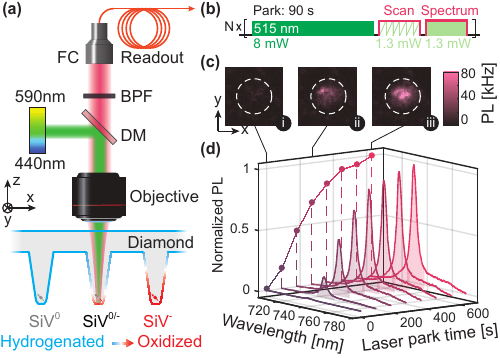}
    \caption{(a) Schematic of the confocal microscope, integrating multiple laser sources (440-595 nm) in the illumination path, separated from the collection path by a dichroic mirror (DM) and flippable bandpass filter (BPF) centered at 740 nm for SiV$^-$ ZPL. The photoluminescence (PL) signal is collected through a fiber coupler (FC) for readout using either a photon-counter or a spectrometer. The objective is mounted on a three-axis piezoelectric scanner enabling scanning individual nanofabricated diamond pillars containing SiV centers. The sample is initially hydrogen-terminated (blue outline) yielding SiV$^0$, with illuminated pillars transitioning to an oxidized surface (red outline), leading to a charge state conversion to SiV$^-$. 
    (b) Measurement sequence for observing charge state dynamics. A continuous-wave (CW) 515 nm laser (8\,mW) illuminates the pillar for 90 s, followed by a low-power (1.3\,mW) confocal x-y scan (zigzag green trace) and PL spectrum acquisition. 
    (c) Three exemplary PL images obtained from x-y confocal scans of a pillar (outlined by a white dashed circle) during laser illumination. 
    (d) PL spectra as a function of accumulated laser park time. The total SiV$^-$ ZPL intensity extracted from each spectrum (shaded area) shows a gradual increase, indicating the recovery of the SiV$^-$ charge state.}
    \label{fig1}
\end{figure}

\textbf{SiV charge state conversion under laser illumination.} 
The sample studied in this work is a commercially available chemical vapor deposition diamond (Element Six, ``EL SC grade'') with a substitutional nitrogen concentration below 5 ppb. 
The sample was implanted with $^{28}$Si at an energy of $80~$keV with a dose of $3\times10^{10}~$ ions/cm$^2$, after which we fabricate parabolic diamond nanopillars\,\cite{Hedrich2020a} with apex diameters of approximately $700~$nm on the diamond surface to enhance excitation and collection efficiencies. 
This results in an average density of three SiV centers per pillar at a mean depth $\langle d\rangle\approx50~$nm according to Stopping Range of Ions in Matter (SRIM) simulations. 
%Details of the sample preparation and characterization can be found in the relevant discussion for sample B in ref.\,\cite{zuber2023shallow}.
%\sout{An additional benefit of nanopillar geometry is their ability to facilitate the study of local effects on a single SiV or a microensemble of SiV. Since SiVs are created near the surface, the initial charge state of SiV is directly controlled by surface termination.} 
%As shown in \,\cite{zhang_neutral_2023}, the charge state of SiV can be toggled between neutral and negative under hydrogen and oxygen surface termination, respectively.

To assess the optical charge state conversion of SiV from SiV$^0$ to SiV$^-$ from an originally H-terminated sample prepared by hydrogen annealing (SI section-\ref{Hter}), we performed a series of photoluminescence (PL) experiments under ambient conditions, using a home-built multicolor confocal microscope setup shown in Fig.\,\ref{fig1}\,(a). 
Specifically, we employ a measurement sequence (Fig.\,\ref{fig1}\,(b)) where we illuminate an H-terminated pillar with a continuous wave (CW) $515~$nm laser for $90~$s at high power ($8~$mW) to drive the charge state conversion. 
Then, we reduce the laser power to $1.3~$mW to probe the SiV 
%avoid driving the process during a
by first recording a confocal image (PL collection window $733~$nm-$747~$nm) of the pillar, and subsequently obtaining a PL spectrum to monitor the strength of the SiV$^-$ emission. 
The sequence of three exemplary PL images shown in Fig.\,\ref{fig1}\,(c) indicates that the amount of PL emission from the pillar increases with each iteration of this sequence, while the accompanying PL spectra demonstrate that this increase in brightness corresponds to the recovery of the SiV$^-$ charge state, as evidenced by the continuously increasing SiV$^-$ ZPL intensity.

\begin{figure}[t!]
    \centering
    \includegraphics[width=85mm]{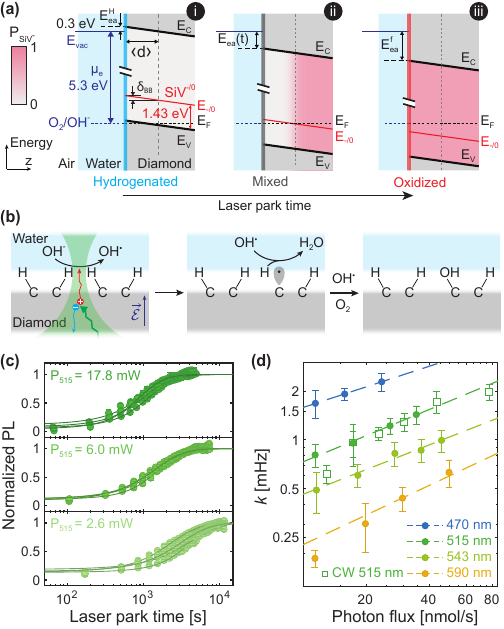}
    \caption{(a) Energy diagram of the diamond at the vicinity of the surface during laser illumination. 
    Stages i-iii qualitatively correspond to the x-y scans shown in Fig.\,\ref{fig1}\,(c), describing the transition from a hydrogen terminated surface to an oxidized one. 
    The vertical dashed line indicates the average depth $\langle d \rangle$ of SiV centers in pillars. 
    The red coloring within the band gap signifies the average population fraction of SiV$^-$ determined by the charge state transition level relative to Fermi level. 
    %The initially hydrogen-terminated surface results a surface Fermi level pinned slightly lower than the valence band maximum (VBM) $E_\text{v}$ due to the electron affinity (EA) of hydrogen terminated diamond in contact with moist air (EA$_{\text{C-H}}\sim$ -0.3 eV) and the electrochemical potentials of the oxygen redox couple in air exposed water ($\mu_e \sim 5.3$ eV), resulting in SiV$^0$ near the surface. With the illumination of the laser, the recovery of the SiV$^-$ suggest a gradual lowering of the charge state transition level caused by an increase of the EA. 
    (b) Proposed photocatalytic reaction for the oxidation of a hydrogen-terminated surface, causing the increase of electron affinity (EA). The photon generated hole migrates to the surface under the electric field $\overrightarrow{\mathcal{E}}$ caused by the band bending and accepted by the oxygen redox couple producing hydroxyl radical $\text{OH}\cdot$. The $\ce{C-H}$ bond on the surface is activated by the hydroxyl radical and can be easily further oxidized.
    (c) The SiV$^-$ ZPL PL intensity (circles) as a function of the $515~$nm laser park time at different powers ($2.6~$mW, $6.0~$mW, $17.8~$mW), where four representative pillars are shown for each power. 
    The time dependent PL increase is fitted with the dynamic charge state model proposed in the text (solid lines). 
    (d) Rate constant $k$ extract from the model as a function of photon flux for four wavelengths. 
    Each data point is averaged over a measurement series on $6-12$ nanopillars under the same wavelength and photon flux. 
    Errorbars show the standard error over the measured ensemble (which is few times higher than the fitting errors). 
    The colored circles represents pulsed laser data at different wavelengths, with power-law fitting (dashed lines).}
    \label{fig2}
\end{figure}

\textbf{Modeling of the charge state conversion dynamics.}  
To obtain further insights into the mechanism underlying the optical charge state conversion process, we continuously record the photon count rate $I_{\text{SiV}^-}$ emitted from the SiV$^-$ ZPL using a $20\,$nm wide bandpass filter centered at $740\,$nm.
Since SiV$^0$ exhibits no emission in this spectral range, the $I_{\text{SiV}^-}$ signal directly reflects the total population of negatively charged SiV centers in the nanopillar. 
%We then determine the charge state population $P_{\text{SiV}^-}\propto I_{\text{SiV}^-}$ through proper normalization, where $P_{\text{SiV}^-} = [\text{SiV}^-]/[\text{SiV}]$ and $[\cdot]$ denotes time-averaged populations.
As energy-level variations induced by SiV depth distributions are negligible compared to room-temperature thermal broadening of the Fermi level, we can treat the total $I_{\text{SiV}^-}$ as proportional to a single effective charge-state population, $P_{\text{SiV}^-} = [\text{SiV}^-]/[\text{SiV}]$, where $[\cdot]$ denotes time-averaged populations (see SI section-\ref{SI_Der}).
Within the adiabatic limit, $P_{\text{SiV}^-}$ follows a Fermi-Dirac distribution\,\cite{defo_charge-state_2023}
\begin{equation}
    P_{\text{SiV}^{-}} = \frac{1}{1+\exp\left[\beta(E_{-/0} - E_F)\right]},
    \label{eq1}
\end{equation}
where $E_{-/0}$ is the adiabatic charge state transition level between SiV$^0$ and SiV$^-$ at the average depth of SiV, $E_F$ is the Fermi level, and $\beta = 1/k_BT$ ($T$ the sample temperature and $k_B$ the Boltzmann constant).
The relevant energy balance (c.f. \,Fig.\,\ref{fig2}\,(a)) can then be expressed as:
\begin{equation}
    E_{-/0} - E_F = \mu_e - E_g -E_{ea} + \Delta - \delta_{BB},
    \label{eq2}
\end{equation}
where $\mu_e$ is the surface electrochemical potential, $E_g$ is the band gap, $E_{ea}$ stands for the electron affinity (EA) defined as $E_{\rm vac} - E_c$, with $E_{\rm vac}$ and $E_c$ the vacuum level and conduction band minimum, respectively, $\Delta$ is the difference between the adiabatic charge state transition level and valence band maximum, and $\delta_{BB}$ is the band bending energy from the surface to the mean depth of SiV.
While the latter has been determined to $\Delta=1.43~$eV for the SiV$^0\leftrightarrow$SiV$^-$ transition\,\cite{gali_ab_2013,thiering_ab_2018}, $\mu_e$ amounts to $5.3~$eV, corresponding to the oxygen redox couple in the thin water layer at the diamond surface (typical for diamond under ambient conditions)\,\cite{chakrapani_charge_2007}.

As a result, the key free parameter determining the average charge state of near-surface SiVs is the EA, $E_{ea}$, which is directly connected to the diamond's surface termination. 
%Combining Eq.\,\eqref{eq1} and Eq.\, \eqref{eq2}, 
Specifically, the continuous tuning of $P_{\text{SiV}^{-}}$ that we observe reflects a gradual reduction of band bending, which arises from a shift of $E_{ea}$ caused by progressive changes in the surface termination, as depicted in Fig.\,\ref{fig2}\,(a). 
The initially H-terminated surface results in a surface Fermi level pinned slightly below the valence band maximum (VBM) $E_\text{v}$, \,\cite{maier_origin_2000, chakrapani_charge_2007} owing to the EA of a H-terminated diamond in contact with moist air of $E_{ea}^{\rm H}\sim-0.3\,$eV and the corresponding surface electrochemical potential $\mu_e$, which results in SiV$^0$ (or SiV$^+$) being the stable charge state for near-surface SiVs. 
With laser illumination, the recovery of 
SiV$^-$ indicates a gradual lowering of the charge state transition level caused by an increase of $E_{ea}$. 
Hence, obtaining the explicit expression of the time evolution $E_{ea}(t)$ is essential to develop a model for the time-dependent SiV$^-$ PL. 
The SiV$^-$ charge state, that we observe to be stabilized after long laser exposure, suggests that the final diamond surface is predominantly oxygen terminated 
(this phenomenon of laser-induced oxidation of H-terminated diamond surfaces has recently been observed and verified by Kelvin probe force microscopy \,\cite{pederson_optical_2024}).
While the photon energy under green illumination is far from sufficient to directly break the $\ce{C-H}$ bond, the overall surface oxidation is an exothermic reaction, suggesting that laser-induced photocatalysis enables a lowering of the activation energy for diamond surface oxidation and thereby accelerates the reaction, as we observe in our experiment \,\cite{rodgers_diamond_2024, holmberg2021photoredox}.

Based on the principle of photocatalysis, we propose the surface photochemical reaction illustrated in Fig.\,\ref{fig2}\,(b) as the mechanism underlying our observations: 
by optical charge cycling of impurities and defects in diamond \,\cite{aslam_photo-induced_2013, zhang_neutral_iti_2023} (s. below), the diamond bulk becomes a continuous source of electrons and holes that are generated in conduction and valence bands, respectively. 
%One necessary ingredient for photocatalysis is a continuous flow of photogenerated charge carriers (electrons or holes) to the surface, which produces free radicals later on.
%Since the photon energy is not sufficient to directly excite electroins across the diamond band gap, we propose that these carriers are generated by charge cycling of impurities and defects in diamond \,\cite{aslam_photo-induced_2013, zhang_neutral_iti_2023} -- a process we will investigate in the following.
The effective electric field caused by the upward band bending in diamond under the H-terminated surface then drives the migration of holes to the diamond surface.
These holes can then be accepted by an oxygen redox pair in the water layer, producing highly reactive hydroxyl radical (OH$\cdot$) \,\cite{parrino_role_2020, chen_unravelling_2024}. 
These hydrogen radicals, in turn, are known to be very efficient in binding protons from alkane chains \,\cite{droege1987hydrogen}, which in the present case activates diamond $\ce{C-H}$ bonds. 
The resulting carbon radical (dangling bond) on the diamond surface is also highly reactive and will readily react with various oxidants in water, finally forming the expected surface oxidation products\,\cite{zhang_photocatalytic_2017}.

The surface photocatalytic reaction we postulated now allows us to obtain an explicit expression for the time evolution of SiV$^-$, where we determine the EA throughout the conversion process as a weighted average between the initial (hydrogenated) and final (oxidized) values, $E_{ea}^{\rm H}$ and $E_{ea}^{f}$, respectively, yielding
\begin{equation}
    E_{ea}(t) = \frac{[\ce{C-H}]_s(t)}{[\text{C}]_s}E_{ea}^{\rm H} + \left( 1- \frac{[\ce{C-H}]_s(t)}{[\text{C}]_s}\right)E_{ea}^{f},
    \label{eq3}
\end{equation}
with $[\ce{C-H}]_s$ the surface concentrations of H-terminated sites and $[\text{C}]_s$ the total concentration of carbon sites at the surface. 
Since the surface activation of a $\ce{C-H}$ bond is a pseudo-first-order reaction (SI section-\ref{SI_Mod}), the time evolution $[\ce{C-H}]_s(t)$ is described by a single exponential decay, where the decay rate $k$ is the reaction rate constant of the surface activation of $\ce{C-H}$ bonds.
Therefore, combining Eqs.\,\eqref{eq1},\,\eqref{eq2}, and \eqref{eq3}, we obtain 
%the SiV$^-$ PL as a function of accumulated laser park time $t$ can be modeled by
\begin{equation}
    I_{\text{SiV}^-}(t) = \frac{D}{1+\exp[A + B\exp(-kt)]}
    \label{eq4}
\end{equation}
where $t$ is the accumulated laser illumination time, $D$ is an asymptote parameter and $k$ must be a function of the photon flux $\Phi$ and illumination wavelength $\lambda$, i.e. $k = k(\Phi,\lambda)$. 
The dimensionless parameters $A$ and $B$ are linear combinations of different energy terms shown in Eq.\,\eqref{eq2}  (SI section-\ref{SI_Mod}), and most importantly, their sum sets the initial surface condition, $A+B=(E_{-/0}^0-E_F)/k_BT$ with $E_{-/0}^0$ the initial position of charge state transition level.
The value of $A+B$ should, therefore, always be positive according to the physical picture presented in Fig.\,\ref{fig2}\,(a). 
As shown in Fig.\,\ref{fig2}\,(c) on the example of $\lambda=515~$nm excitation, Eq.\,\eqref{eq4} provides for excellent fits (solid lines) to experimentally measured traces $I_{\text{SiV}^-}(t)$, that allow for quantitative extraction of $k$ as a function of experimental parameters.

\textbf{Photon flux dependence of the reaction rate $k$.} 
To gain further insight into the photochemistry underlying the optical charge state conversion of shallow SiVs, we determine the reaction rate $k$ for a range of SiVs as a function of photon flux and irradiation wavelength, using a supercontinuum laser (NKT, SuperK FIANIUM)  as a tunable source for optical charge conversion.
For each power and wavelength, we obtain $k$ by averaging the fitting results from several (typically $6-12$) pillars, to mitigate slight inhomogeneities in $k$ caused by the random positioning of the impurities responsible for hole generation.
Figure\,\ref{fig2}\,(d) shows the resulting power- and wavelength-dependence of $k$, where the log-log plot reveals a power law scaling $k \propto \Phi^\beta$ for all investigated wavelengths.
We find $\beta\approx0.5$, which is characteristic for a photocatalytic reaction rate in the ``intermediate optical power regime'' \,\cite{buglioni_technological_2022}, where the impurities responsible for the electron-hole pair generation are driven above saturation, such that the two-photon electron-hole generation process scales linearly with $\Phi$\,\cite{aslam_photo-induced_2013,siyushev2019photoelectrical}, and most of the generated electron-hole pairs undergo recombination, while a small fraction of the charge carriers participate in the photocatalytic reaction (SI section-\ref{SI_Mod}).
We note that while our supercontinuum laser provides for pulsed laser excitation, in our regime of operation, we observe no difference in $k$ between pulsed and CW excitation for the same average output powers.
We explicitly verify this for the case of $515~$nm excitation, where within error bars, we observe near-identical scalings of $k$ with photon flux, as shown in Fig.\,\ref{fig2}\,(d), for pulsed (green circles) and CW (green squares) excitation.

 \begin{figure}[t!]
    \centering
    \includegraphics[width=83mm]{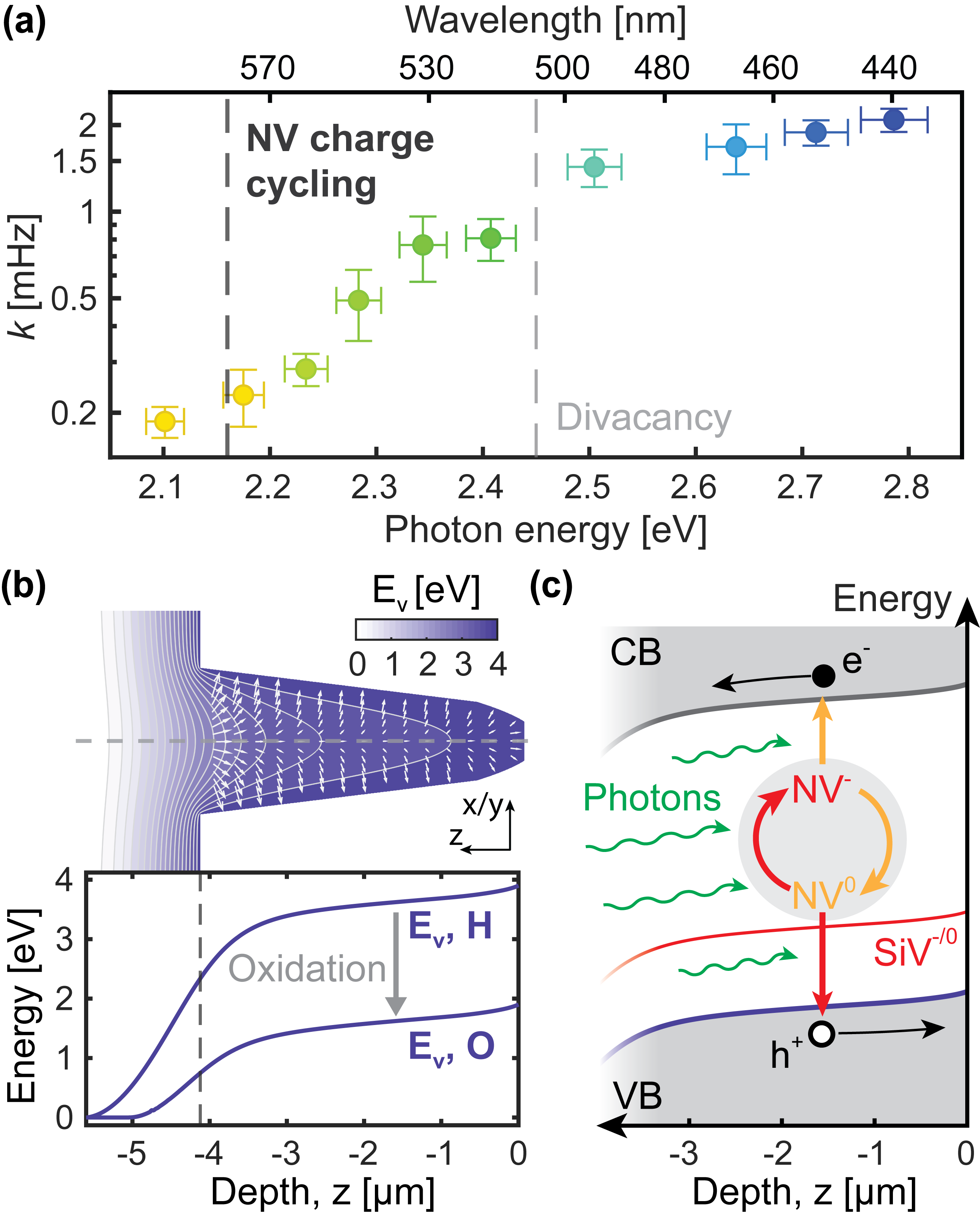}
    \caption{(a) Photon energy dependency of the rate constant $k$, measured for different wavelengths under the same photon flux of $11~$nmol/s. 
    The charge cycling threshold energies of the NV ($2.16~$eV) and the divacany ($2.45~$eV) are indicated with dashed lines and mark the onset of prominent steps in $k$. 
    The errorbars in $k$ represents the standard error over the measured ensemble and the errorbar in photon energy is given by the minimum linewidth ($\pm5$\,nm) of the filter of the supercontinuum laser.
    (b) Calculated valence band maximum ($E_\text{v}$) with respect to its bulk limit in the pillar geometry.
    The 2D map of $E_\text{v}$ is calculated for a H-terminated surface and the arrows indicates the relative strength and direction of the local electric field induced by the band bending.
    Profiles along a line-cut at the pillar center are depicted for a H- and O-terminated surface (bottom panel). 
    %These results show that upwards band bending (in type-IIa diamond) caused by the surface termination can extend along the entire pillar length. 
    (c) Mechanism of the electron-hole pair generation by charge cycling of deep NVs and charge carrier migration under the electric field associated with the band bending in diamond.}
    \label{fig3}
\end{figure}

%\subsection{Photon energy dependency of conversion rate}

\textbf{Photon energy dependence of the reaction rate $k$.} 
To identify the source of photo-generated charge carriers driving photocatalysis, we next investigate the wavelength (photon energy) dependence of the reaction rate $k$ at fixed photon flux. 
To maximize spectral coverage, we extend the wavelengths in this study down to $445~$nm, and set the photon flux to $11\,$nmol/s, corresponding to the maximal available power ($\sim3\,$mW) at $445~$nm.
For each wavelength, we average $k$, as extracted from fitting SiV$^-$ PL time traces, over $6$-$12$ pillars. 
The resulting average value of $k$, plotted versus photon energy on a semi-logarithmic scale, is shown in Fig.\,\ref{fig3}\,(a). 
We clearly observe a monotonic increase of $k$ with photon energy, with a first marked step-increase initiating beyond $\sim2.2\,$eV, that we attribute to NV  charge-cycling, which becomes effective for photon energies higher than the NV$^0$ ZPL at $2.16\,$eV \,\cite{lozovoi2021optical}. 
Instead of a saturation of $k$, expected if NV centers were the sole contributor to hole generation, a second, less pronounced step is observed around $2.45\,$eV, which we tentatively attribute to additional hole generation from charge cycling of divacancies in the vicinity of the SiVs under investigation\,\cite{gorlitz_coherence_2022, rieger_fast_2024}. 

Given the low impurity density in electronic-grade diamond and the absence of implanted NV centers in our sample, the hole generation by NV charge cycling that we postulate must be related to NVs at a sizable distance from the SiVs under investigation.  
We now estimate the spatial extent over which defects could thereby effectively contribute to photocatalysis by determining the band bending profile within and near the diamond pillar by numerically solving for the Poisson equation with appropriate boundary conditions (SI section-\ref{SI_BB}). 
In Fig.\,\ref{fig3}\,(b) (top panel), we illustrate the resulting VBM profile for a hydrogen-terminated surface, alongside the induced electric field within the pillar. 
The profile along the pillar symmetry axis (Fig.\,\ref{fig3}\,(b), bottom panel) indicates that an effective electric field up to $10^5\,$V/m builds up within a few microns from the pillar's apex -- an electric field magnitude that is sufficient for efficient charge carrier migration in diamond\,\cite{lozovoi_detection_2023}. 
Moreover, these simulations show that a nonzero electric field persists throughout the surface photocatalytic reaction process that gradually ``flattens'' the bands, as indicated by results for an oxygen-terminated surface. 
This analysis suggests that even remote, deep, native NVs in the pillar, or even the surrounding bulk can effectively contribute to the surface oxidation, as demonstrated in Fig.\,\ref{fig3}\,(c).

The presence of such deep, native NV centers is verified by the consistent occurrence of characteristic NV$^-$ PL spectral signatures in the background of the SiV PL spectra\,(SI section-\ref{SI_NVPL}). 
The key role of NV centers in driving the photocatalytic process is further supported by our observation of significantly enhanced rate constants $k$ when repeating our experiments under nominally identical conditions (at $532~$nm excitation) on nanopillars that are implanted with shallow NV centers\, (SI section-\ref{SI_NVconv}). 
This enhancement arises from the increased fraction of photo-generated holes contributing to the surface photocatalytic oxidation, as shallow NVs, being closer to the surface, are more efficient at delivering carriers than native, deep NVs.
%While divacancies remain spectroscopically undetectable, their role in the photocatalytic process in these control experiments is supported by an additional increase in the reaction rates $k$ comparing 480\,nm to 532\,nm laser illumination in pillars containing shallow NVs\,(SI section-\ref{SI_NVconv}).
%Under H-termination, these NV centers are initially stabilized to NV$^+$, which completely suppresses the possibility of optical charge cycling and hole generation under our experimental conditions. 
%Yet, we observe optical charge state conversion of these NVs with distinctively enhanced reaction rates $k$ when comparing blue to green laser illumination\,(SI section-\ref{SI_NVconv}).

%\subsection{Charge state and surface under cryogenic environment}

 \begin{figure}[t!]
    \centering
    \includegraphics[width=83mm]{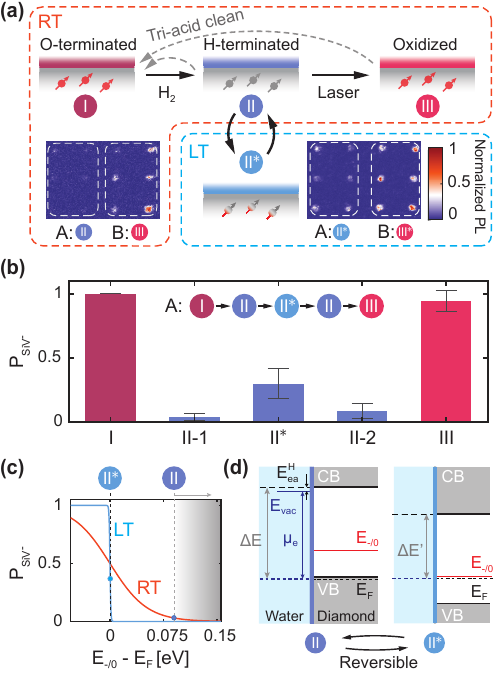}
    \caption{(a) Schematic of surface configurations at room temperature (RT) and cryogenic temperature (LT) used to study the influence of cryogenic environment on the charge state of SiV under a H-terminated surface. 
    State-I represents an O-terminated surface at (RT) prepared by a tri-acid clean where the SiVs are considered to be fully in the negative charge state. 
    State-II represents a H-terminated surface at RT, and state-II* is the same surface termination cooled to cryogenic temperature. 
    State-III denotes an oxidized H-terminated surface at RT after laser illumination. 
    Confocal scans show SiV$^-$ PL intensity for two groups of six nanopillars each: Group A remains unexposed to laser after H-termination, while Group B serves as a reference, having undergone laser-induced oxidation.
    (b) The population fraction of SiV$^-$ $P_{\text{SiV}^-}$ at each state in the sequence from I, II, II*, II to III, measured as the average of SiV$^-$ PL intensity for each pillar in Group A normalized to state-I.
    (c) Plot of $P_{\text{SiV}^-}$ as a function of Fermi level relative to the charge state transition level $E_{-/0}$ for RT and LT conditions shows the increase of the SiV$^-$ population at LT is correlated to a Fermi level shift.
    (d) Possible energy band diagrams for H-terminated surface at RT (state-II) and LT (state-II*), illustrating a shift in surface Fermi level due to the change in the difference of electrochemical potential and EA between RT ($\Delta E$) and LT ($\Delta E'$)conditions.
    }
    \label{fig4}
\end{figure}

\textbf{Surface state and near-surface SiV charge state under cryogenic conditions.} 
%Finally, we address how the tunability of the surface termination via our laser oxidation method may optimize the charge state of an emitter, particularly SiV$^0$. Under ambient conditions and hydrogen termination the diamond Fermi level is located near the transition between neutral and positive charge states for shallow SiVs (see Fig.\,\ref{fig2}a). Consequently, either the positively charged SiV is stabilised, or the SiV charge state hops between neutral and positively charged\,\cite{gali_ab_2013, pederson_optical_2024}. The ability to tune the charge state transition level such that SiV$^0$ is the stable charge state would be beneficial for further explorations of this elusive, yet highly attractive color center\,\cite{rose_observation_2018}. However, SiV$^0$ needs to be kept in cryogenic environments to maintain favorable spin and optical properties, which can significantly affect the surface state. Recent work also suggests that it may suffer from a poor signal-to-noise ratio at cryogenic temperatures\,\cite{zhang_neutral_2023}. 
%At ambient conditions, hydrogen termination likely positions the Fermi level near the transition between neutral and positive charge states. In this case, being able to bring down the charge state transition level would indeed be beneficial for achieving a perfectly stabilized SiV$^0$. 
%However, the operational conditions for SiV$^0$, which, typically in a cryogenic environment to ensure good spin and optical properties, can significantly alter the surface state. 
Finally, we investigate the impact of the diamond surface chemistry on the charge state of shallow SiVs under cryogenic conditions -- the typical operating conditions for both SiV$^0$ and SiV$^-$ -- 
to explore the potential of using H-termination together with the tunability of laser oxidation to optimize SiV$^0$ charge state stability.
Figure\,\ref{fig4}\,(a) presents an overview of the surface terminations we investigate under ambient (``RT'') and cryogenic (``LT''; temperature $T\lesssim7~$K and pressure $p<10^{-5}~$mBar) conditions and how they convert into each other. 
We initially prepare the sample in its O-terminated state (``state-I'' in Fig.\,\ref{fig4}\,(a)) by tri-acid cleaning, and subsequently perform H-termination (``state-II'' in Fig.\,\ref{fig4}\,(a)), as described earlier.
We assess how the cryogenic conditions affect SiVs under an H-terminated surface, on a set of six pillars (denoted as ``group A''), by comparing the ZPL intensity $I_{\text{SiV}^-}$ between RT and LT conditions. 
To obtain an estimate for $P_{\text{SiV}^-}$, for each pillar, we normalize the measured values of $I_{\text{SiV}^-}$ in each state by the value for $I_{\text{SiV}^-}$ we obtained in state-I where SiVs are considered to be fully in their negative state, under identical excitation conditions.
In addition, we prepare a second group (``Group B'') of six further pillars in an oxidized state by laser surface oxidation to act as a reference group on the same sample during our measurements at RT and LT.

%As expected, at RT we observe that the 
After cooling down the sample, we observe that $I_{\text{SiV}^-}$ for pillars in group A, that were completely quenched in state II, partially recovers, yielding $P_{\text{SiV}^-}=0.30\pm0.12$ under LT conditions (``state-II*'' in Fig.\,\ref{fig4}\,(a)), as shown in the confocal scans and quantitative analysis in Fig.\,\ref{fig4}\,(a), and (b), respectively. 
%In comparison to reference Group B of SiV previously prepared with an oxidized surface, the pillars in Group A at state-II* become visible, whereas in state-II, they remain completely dark. 
For comparison, $I_{\text{SiV}^-}$ in reference group B only shows a mild increase in $I_{\text{SiV}^-}$, demonstrating that the partial recovery of $I_{\text{SiV}^-}$ we observed for group A in state-II* is indeed related to the transition to LT conditions and not an experimental artifact.
Lastly, the observed process is completely reversible, in that warming up the sample back to state-II, again yields a near-complete suppression of $I_{\text{SiV}^-}$, and subsequent laser oxidation of the pillars in group A fully restores $P_{\text{SiV}^-}$ back to nearly one (Fig.\,\ref{fig4}\,(b)).

%The quantitative analysis of this phenomenon, revealed by the SiV charge state population, is depicted in Fig.\,\ref{fig4}\,(b). 
%The results align with expectations for stage II, where almost no SiV$^-$ PL is observed, and for state-III, where the SiV$^-$ PL recovers to the same level as an oxygen-terminated surface, indicating fully recovered SiV$^-$. 
%More interestingly, this result suggests that the SiV$^-$ population is partially recovered to $30\%\pm12\%$ in H-terminated diamond at cryogenic conditions.
%Moreover, this change in brightness is reversible: when the sample is warmed to RT conditions, SiV$^-$ becomes dark again. 

Our findings suggest that under cryogenic conditions, the band bending induced by H-termination becomes significantly less effective than at room temperature, shifting the charge-state transition level $E_{-/0}$ very close to $E_F$, which indicates a shift of $E_{-/0}-E_F$ of at least $1.4$\,eV (See Fig.\,\ref{fig4}\,(d)) between RT and LT conditions.
Based on our earlier discussion (c.f. Eq.\,\eqref{eq1}), this shift must be caused by a corresponding shift in the difference between electrochemical potentials $\mu_e$ and $E_{ea}$ due to modifications of surface properties when going from RT to LT conditions.
We tentatively assign this modification to either the removal or freezing out of the water layer in cryogenic conditions, both of which will have a marked impact on $\mu_e$ and $E_{ea}$.
To elucidate the exact origin of this effect and corroborate our hypothesis, further experimental investigations, such as surface conductivity measurements at LT\,\cite{nebel_low_2002} could be employed, but are beyond the scope of this work.

\section{Conclusion}

In summary, we demonstrated a continuous optical charge state tuning method for shallow SiV centers in diamond nanopillars via a well-controlled laser-induced surface oxidation process of H-terminated diamond. 
Our approach allows on-demand formation of an arbitrary intermediate surface configuration between H- and O-terminated surfaces, offering a powerful tool for precise charge state control across varying color-center depths. 
By leveraging the SiV$^-$ PL intensity as an estimator for the SiV's probability to occupy the negative charge state, 
we gained a microscopic understanding of this laser-induced surface oxidation by modeling its time dependency and quantitatively extracting the relevant rate constant for surface oxidation.
Our systematic study of the rate constant as a function of photon flux and photon energy revealed the underlying photocatalytic processes and allowed us to identify NV center and divacancy charge cycling as the main sources of the photogenerated carriers driving the surface photocatalysis. 
Key to this process in ultrapure diamond is the use of diamond nanostructures that allow for efficient collection of these carriers towards the surface, thereby boosting surface oxidation timescales by $1-2$ orders of magnitude compared to previous studies\,\cite{rodgers_diamond_2024}.

Looking forward, the growing interest in both SiV$^0$ and SiV$^-$ for diverse quantum applications highlights the relevance of our findings for device design and operational protocols, aiming at stable charge states and advanced charge state control.
Although we propose a plausible surface chemical reaction leading primarily to hydroxyl termination after laser oxidation, further experimental validation, such as XPS and nanoscale NMR, would be helpful to further determine the precise, resulting surface configuration. 
A deeper understanding of these reactions could pave the way for novel approaches to local functionalization of the surface of diamond nanostructures\,\cite{rodgers_diamond_2024}.
Since optical tuning of the charge state arises from surface chemical modifications of diamond, this effect is applicable to other near-surface color centers beyond SiV.
Moreover, the efficient tunability of diamond's surface termination that we demonstrated, apart from the application in charge state stabilization, may also contribute to reduced spin noise\,\cite{ryan_impact_2018,gulka_surface_2024}, suppressed photobleaching and blinking of near-surface color centers\,\cite{kaviani_proper_2014, stacey_evidence_2019}, and thereby be of broad importance for future quantum technology applications of shallow color centers in diamond nanostructures.

Finally, we have investigated the potential for enhancing the charge state stability of SiV$^0$ through our optical charge state tuning method. Unexpectedly, a partially recovered SiV$^-$ charge state was detected in H-terminated diamond at cryogenic temperature. 
This observation suggests that the charge state instability in such an environment may stem from charge state hopping between SiV$^-$ and SiV$^0$.
While the underlying mechanism for the Fermi level shift remains to be fully uncovered, our findings offer valuable insight into the challenges associated with charge stabilization of SiV$^0$ prepared under H-termination at LT\,\cite{zhang_neutral_2023}. Understanding these results may also inform future strategies for mitigating these charge state instabilities\,\cite{neethirajan_controlled_2023}, paving the way towards deterministic initialization of stable and bright SiV$^0$ centers. 

\section{acknowledgments}
\begin{acknowledgments}
We gratefully acknowledge Christoph Becher and Nathalie de Leon for fruitful discussions, as well as Brendan J. Shields for the help in fabricating the diamond nanopillar sample. 
We acknowledge financial support through the QuantERA project ``sensExtreme'' (Grant No. 205573), from the Swiss Nanoscience Institute, and through the Swiss NSF Project Grant No. 188521.
A.C. acknowledges financial support from the Quantum Science and Technologies at the European Campus (QUSTEC) project of the European Union’s Horizon 2020 research and innovation program under the Marie Skłodowska-Curie grant agreement No.\,847471.
\end{acknowledgments}

\bibliographystyle{apsrev4-2}
\bibliography{Main_Optical_charge_state_conv.bib}

\end{document}

\title{Supplementary Information for: Photocatalytic Control of Diamond Color Center Charge States via Surface Oxidation}
\author{Minghao Li}
\email{minghao.li@unibas.ch}
\thanks{These authors contributed equally.} 
\affiliation{Department of Physics, University of Basel, CH-4056 Basel, Switzerland}
\author{Josh A. Zuber}
\thanks{These authors contributed equally.} 
\affiliation{Department of Physics, University of Basel, CH-4056 Basel, Switzerland}
\affiliation{Swiss Nanoscience Institute, University of Basel, CH-4056 Basel, Switzerland}
\author{Marina Obramenko}
\affiliation{Department of Physics, University of Basel, CH-4056 Basel, Switzerland}
\affiliation{Swiss Nanoscience Institute, University of Basel, CH-4056 Basel, Switzerland}
\author{Patrik Tognina}
\affiliation{Department of Physics, University of Basel, CH-4056 Basel, Switzerland}
\author{Andrea Corazza}
\affiliation{Department of Physics, University of Basel, CH-4056 Basel, Switzerland}
\author{Marietta Batzer}
\affiliation{Department of Physics, University of Basel, CH-4056 Basel, Switzerland}
\author{Marcel.li Grimau Puigibert}
\affiliation{Department of Physics, University of Basel, CH-4056 Basel, Switzerland}
\author{Jodok Happacher}
\affiliation{Department of Physics, University of Basel, CH-4056 Basel, Switzerland}
\author{Patrick Maletinsky}
\email{patrick.maletinsky@unibas.ch}
\affiliation{Department of Physics, University of Basel, CH-4056 Basel, Switzerland}
\affiliation{Swiss Nanoscience Institute, University of Basel, CH-4056 Basel, Switzerland}

\maketitle

\section{Optical setup}

The optical setup used to perform optical charge-state conversion of SiV centers via surface photocatalytic oxidation is a home-built multicolor confocal microscope setup operating under ambient conditions, sketched in Fig.\,\ref{fig1}\,(a).
Excitation and collection paths are fiber-coupled (single-mode fibers) and focused onto the sample using a 0.8 NA microscope objective (Olympus LMPLFLN 100$\times$).
For excitation, a 515\,nm diode laser (Cobolt 06-MLD) and a supercontinuum pulsed laser (SuperK Fianium FIU-15) coupled to a SuperK VARIA tunable filter (minimum linewidth 10\,nm) are used.
In the collection path, the photoluminescence (PL) is detected either by an avalanche photodiode (silicon-based single-photon counting modules, SPCM, Excelitas AQRH-33) for time-resolved measurements, or recorded on a CCD camera (PIXIS100) coupled to a monochromator (Princeton Instruments SP-2500) for spectral analysis. 
The selective collection of SiV$^-$ zero-phonon line (ZPL) is filtered by a bandpass filter centered at 740\,nm with 20\,nm full width at half maximum (FWHM) (Semrock FBP01-740/13-12.5).
For the temperature-dependent charge-state measurements (presented in Fig.\,\ref{fig4}), the sample is housed in a closed-cycle variable temperature cryostat (attocube attoDRY800, base temperature at sample $\sim$ 6.5\,K).
Optical access is provided by a similar home-built confocal microscope equipped with the same objective and excitation/detection components.

%%%%%%%%%%%%%
\section{Sample preparation and fabrication}

We prepare the sample starting from a commercially available electronic grade diamond (Element Six), sliced and polished into a \qty{50}{\um}-thick platelet. 
We then implant $^{28}$Si$^+$ ions at 80\,keV, with a dose of $3\times10^{10}$\,ions/cm$^2$, at an angle of 7$^{\circ}$. The resulting emitters have an average depth of 50\,nm, as estimated by Stopping Range of Ions in Matter (SRIM) simulations.
To form SiV centers, we anneal the sample in a home-built vacuum oven with steps at 400\,$^{\circ}$C, 800\,$^{\circ}$C and 1300\,$^{\circ}$C and durations of 4\,h, 11\,h and 2\,h, respectively.

\begin{figure}[h]
    \centering
    \includegraphics[width=150mm]{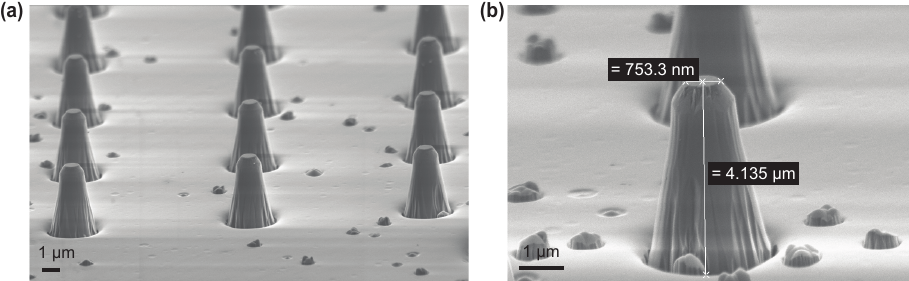}
    \caption{(a) A SEM overview of part of the write field of the diamond nanopillar sample where the data of our optical charge state conversion are acquired. 
    (b) A zoom-in SEM image on one pillar with the typical size of the pillar in this write field.}
    \label{figS1}
\end{figure}

Next, we fabricate diamond nanopillars with apex diameters of 700\,nm to enhance the excitation and collection efficiency (c.f. Fig.\,\ref{figS1}\,(a)),  following the process described in \cite{Hedrich2020a}.
To restore and preserve optical coherence of the SiV in the nanopillar after fabrication, we repeat the same annealing procedure a second time. The final structures contain, on average, three SiV centers per pillar, at a mean depth of 50\,nm.

%%%%%%%%%%%%%
\section{Preparation of surface terminations}
\label{Hter}
The initially oxygen-terminated diamond surface in this work is prepared by the tri-acid clean, where the diamond is submerged in a heated and refluxing mixture (1:1:1) of nitric, sulfuric and perchloric acid. The hydrogen termination of the diamond sample is realized by hydrogen annealing, where the sample is annealed in a pure H$_2$ atmosphere at 750$\degree$ for 6 hours in a tubular furnace\,\cite{fizzotti2007diamond, seshan2013hydrogen}.
The detailed characterization including XPS and AFM of thusly formed surface termination is presented in \,\cite{zhang_neutral_2023}.
All optical charge state conversion measurements are taken right after a freshly H-terminated sample to avoid the impact of slow degradation (oxidation caused by oxygen in the air) of the H-terminated surface in the atmosphere.

%%%%%%%%%%%%%
%\section{Scanning electron microscopy (SEM) image of the diamond nanopillar sample}

%%%%%%%%%%%%%
\section{Additional observation of the optical charge state conversion}
%%%%%%
\label{SI_add}
\subsection{PL dynamics in dark}

\begin{figure}[h]
    \centering
    \includegraphics[width=135mm]{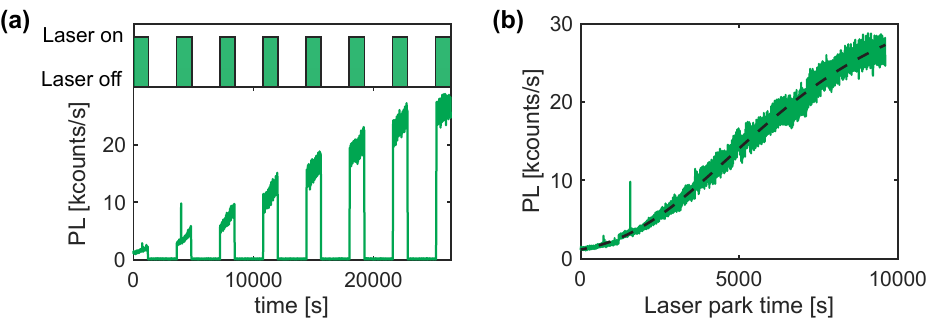}
    \caption{(a) Time evolution of SiV ZPL photoluminescence (PL) intensity (lower panel) under periodically gated 515\,nm laser illumination on an initially H-terminated diamond nanopillar. The illumination sequence is illustrated in the upper panel: each cycle consists of 1200\,s of laser on at 1.5\,mW, followed by 2400\,s of laser off.
    (b) Reconstructed PL time trace showing only the laser-on segments from (a), plotted continuously as a function of cumulative laser-on (or "laser park") time, fitted with the model proposed in the main text (black dashed line).}
    \label{figS31}
\end{figure}

To gain a better understanding of the origin of the observed optical charge state conversion, we investigated the photoluminescence (PL) dynamics of SiV centers when the laser is periodically switched off during the conversion process.  
To this end, we design a pulsed laser sequence, as illustrated in the upper panel of Fig. \ref{figS31} (a).  The corresponding time evolution of the SiV zero-phonon line (ZPL) PL is shown in the lower panel of Fig. \ref{figS31} (a). 
The results clearly show that the PL intensity remains unchanged during the laser-off period and resumes its increase immediately once the laser is turned back on. 
To better analyze the PL evolution dynamics, we extracted and concatenated only the laser-on segments of the trace. The resulting "articulated" PL curve is shown in Fig. \ref{figS31} (b). 
The reconstructed trace is indistinguishable from a PL trace acquired under continuous illumination and can be well fitted by the same model of PL dynamics proposed in the main text, with the extracted reaction rate constant $k = 0.29\pm0.02\,$ mHz, consistent with the expected value for 515\,nm illumination at 1.5\,mW, based on the photon flux dependence. 
These results confirm the adiabatic nature of the charge state conversion, where a well-defined charge state is stabilized at each moment. This feature enables fine control over the charge state through laser illumination, corresponding to a continuously tunable surface termination. 
In principle, this study shows the ability of preparing arbitrary intermediate surface configurations between the initial and final surface terminations using our approach of optical tuning of charge state.

%%%%%%ANDREA CORAZZA%%%%%%
\subsection{SiV charge state conversion observed in thin diamond membranes}
The membrane samples were fabricated from \qtyproduct[product-units=power]{4.5 x 4.5}{\mm}, \qty{500}{\um}-thick electronic-grade (100) diamond plates (Element Six; [N\textsubscript{s}]\,<\,5 ppb, [B]\,<\,1 ppb) grown by chemical vapor deposition. These plates were laser-sliced to \qty{\sim 50}{\um} thickness and double-side polished to $R_q<1\,$nm by Almax EasyLab (Belgium). Front-side patterning of \qtyproduct[product-units=power]{20 x 20}{\um} diamond platelets, each tethered to the holding frame by a \qty{1}{\um} wide bridge, was performed via electron-beam lithography (Zeiss Supra with Raith Elphy) followed by O\(_2\) ICP-RIE etching according to Ref.~\cite{Appel2016}. 
A deep etch on the backside using Ar / Cl\(_2\) and O\(_2\) plasmas and a bulk quartz mask~\cite{Appel2016,Challier2018}, then released free-standing platelets with final thicknesses of \qtyrange{500}{900}{\nm}.

For this study, two samples were fabricated:
\begin{itemize}
  \item \textbf{M1:} Implanted first with \textsuperscript{14}N (55 keV, $2\times10^{9}$ ions/cm\(^2\) at 7$^\circ$), then with \textsuperscript{29}Si (80 keV, $1\times10^{10}$ ions/cm\(^2\) at 7$^\circ$) by Innovion (now Coherent, USA), yielding mean depths of 66\,nm (NV) and 50\,nm (SiV) estimated by SRIM.
  \item \textbf{M2:} Implanted only with \textsuperscript{29}Si at 140 keV and 7$^\circ$ in regions of $3.02\times10^{10}$ and $7\times10^{9}$ ions/cm\(^2\) (CuttingEdge Ions, USA), resulting in a mean depth of \qty{100}{\nm} (SRIM).
\end{itemize}

\begin{figure}[t]
    \centering
    \includegraphics[width=150mm]{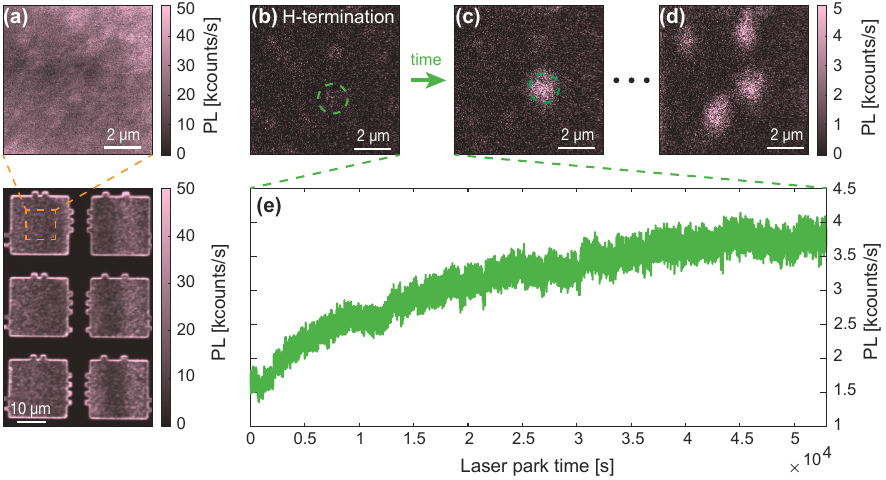}
  \caption{SiV charge–state control in membrane sample M1. 
    (a) Wide‐field confocal fluorescence image of the patterned membrane sample, showing six \qtyproduct[product-units=power]{20 x 20}{\um} platelets (700–800 nm thick) after oxygen termination (tri‐acid clean). PL was collected over 600–800 nm. The inset shows the investigated area on sample M1 for the conversion study.
    (b) Confocal image of the investigated region after hydrogen termination, with PL filtered to 700–750 nm; the characteristic SiV$^{-}$ emission is quenched. 
    (c) Localized recovery of SiV$^{-}$ emission by illumination with a \qty{515}{\nm}, 20 mW laser over the area in (b). 
    (d) SiV conversion at four discrete membrane locations using the same 515 nm, 20 mW laser illumination. 
    (e) Representative time trace of SiV$^{-}$ PL intensity under continuous 515 nm (20 mW) exposure, illustrating the charge‐state conversion dynamics.}
  \label{figS32}
\end{figure}

Following ion implantation, SiV centers were activated by high-temperature vacuum annealing, as detailed in~\cite{zuber2023shallow}. Fig.\,\ref{figS32}\,(a) shows a representative confocal fluorescence scan of an oxygen-terminated membrane (after tri-acid clean), with SiV$^{-}$ emission collected from 600\,nm to 800\,nm. The inset highlights the 700\,nm-thick platelet on sample M1 used for the charge-state conversion study. After hydrogen termination, SiV$^{-}$ photoluminescence is effectively quenched (Fig.~\ref{figS32}(b), PL filtered to 700–750\,nm). This quenching behavior is consistent between the ensemble and single-center regions in sample M2.  

By parking a 515\,nm laser at 10–20\,mW in a fixed spot for several hours, we observe a gradual recovery of SiV$^{-}$ PL (Fig.~\ref{figS32}(c)), with the time dependence of SiV$^{-}$ fluorescence shown in Fig.~\ref{figS32}(e) (20\,mW, sample M1). 
The restored SiV$^{-}$ emission remains stable over time after prolonged illumination.
Since this PL dynamics is observed in a flat and thin diamond membrane which is not confined as a nanopillar, due to the diffusion of charge carriers, this effect becomes non-local, and the collected PL intensity is no longer from a fixed number of emitters, making our model invalid here to fit the dynamics. 
However, we can still see that the time scale to reach the PL saturation is much longer than the one observed in our pillar, indicating the enhancement of the photocatalysis rate by our nanostructure compared to an unstructured surface.
Moreover, we activate SiV$^{-}$ at multiple spots on the same membrane (Fig.~\ref{figS32}(d)) by the prolonged 515\,nm laser illumination.

%%%%%%%%%%%%%
\section{Model for the photocatalytic surface oxidation}
\label{SI_Mod}
In this section, we propose plausible surface chemical reactions with the associated rate constant for each step. Here, we note that the actual surface photocatalytic reaction can be extremely complex, which forms an entire field of study\,\cite{raymakers_diamond_2019, holmberg2021photoredox}. However, based on knowledge of the well-known chemical potential of the diamond surface adsorbate along with the charge transfer and photocatalysis mechanism\,\cite{chakrapani_charge_2007}, we are able to build a plausible rate equation model combining photocharge carrier generation and surface reactions to elucidate the 0.5 scaling order for the photon flux dependency of the reaction rate constant. The corresponding regimes of such scaling order allow us to reproduce the order of magnitude of the resulting macroscopic C-H bond activation reaction rate.

%%%%%%
\subsection{Rate equation model for NV charge cycling}

To study the photon flux dependence of the total surface oxidation reaction rate, we start by looking at the dynamics of the generation of electron-hole pairs in diamond, as a continuous flow of photogenerated electron-hole pairs is essential for the production of free radicals that accelerate the reaction. 
In our experiment, the photon energy is not sufficient to directly excite through the diamond band gap ($\sim 5.5$\,eV). However, with the help of the charge cycle of various defects lying within the diamond band gap, the generation of charge carriers could be performed through a multiple-step excitation.
Here, for simplification without loss of generality, we take the NV center charge cycle, which has been shown to be the main source of electron-hole pairs in the main text, as an example to build our rate equations.

The NV charge cycle, starting with a negatively charged NV in its ground state, can be decomposed into the following steps:
\begin{itemize}
    \item [1.] The NV$^-$ gets excited by a photon to its excited state NV$^{-,*}$
    \begin{equation}
    \text{NV}^- + h\nu \xrightarrow[]{k_1} \text{NV}^{-,*}
    \end{equation}
    \item [2.] Once the NV$^-$ is excited to the excited state, two paths are possible for the population. It can either relaxed to the ground state:
    \begin{equation}
    \text{NV}^{-,*} \xrightarrow[]{\Gamma_1} \text{NV}^-
    \end{equation}
    or be ionized by exciting an electron to the conduction band.
    \begin{equation}
    \text{NV}^{-,*} + h\nu \xrightarrow[]{k_2} \text{NV}^0 + \text{e}^-
    \end{equation}
    \item [3.] Once the NV$^-$ is ionized, it can be excited again to the excited state of NV$^0$
    \begin{equation}
    \text{NV}^0 + h\nu \xrightarrow[]{k_3} \text{NV}^{0,*}
    \end{equation}
    \item [4.] Similarly, when the NV$^0$ is in its excited state, it can either relax to its ground state
    \begin{equation}
    \text{NV}^{0,*} \xrightarrow[]{\Gamma_0} \text{NV}^0
    \end{equation}
    or be recombined with one electron excited from the valence band, leaving a hole in the valence band and return to NV$^-$ charge state.
    \begin{equation}
    \text{NV}^{0,*} + h\nu \xrightarrow[]{k_4} \text{NV}^- + \text{h}^+
    \end{equation}
\end{itemize}

Using these dynamics and associated rate constants, we can write rate equations for the population of ground- and excited-state NV in different charge states ($[\cdot]$ denotes the population, and $I$ is the light intensity proportional to the photon flux $\Phi$):
\begin{equation}
\frac{d}{dt}[\text{NV}^{-,*}] = k_1I [\text{NV}^-] - \Gamma_1[\text{NV}^{-,*}] + k_2I[\text{NV}^{-,*}]
\end{equation}
\begin{equation}
\frac{d}{dt}[\text{NV}^{0}] = k_2I [\text{NV}^{-,*}] - k_3I[\text{NV}^0] + \Gamma_0[\text{NV}^{0,*}]
\end{equation}
\begin{equation}
\frac{d}{dt}[\text{NV}^{0,*}] = k_3I [\text{NV}^0] - \Gamma_0[\text{NV}^{0,*}] - k_4I[\text{NV}^{0,*}]
\end{equation}
\begin{equation}
\frac{d}{dt}[\text{NV}^{-}] = k_4I [\text{NV}^{0,*}]-k_1I[\text{NV}^-]+\Gamma_1[\text{NV}^{-,*}]
\end{equation}
Since we have continuous illumination in our experiment, we consider here only the steady state, where
\begin{equation}
\frac{d}{dt}[\text{NV}^{-,*}] = \frac{d}{dt}[\text{NV}^{0}] =\frac{d}{dt}[\text{NV}^{0,*}] =\frac{d}{dt}[\text{NV}^{-}] = 0.
\end{equation}
We can thus obtain the relations that determine the population of the ground and the excited state of NV$^0$ and NV$^-$:
\begin{equation}
k_1I[\text{NV}^-] = (\Gamma_1 + k_2I) [\text{NV}^{-,*}]
\label{eqNV1}
\end{equation}
\begin{equation}
k_3I[\text{NV}^0] = (\Gamma_0 + k_4I) [\text{NV}^{0,*}]
\end{equation}
\begin{equation}
\frac{k_1 k_2}{\Gamma_1 + k_2I}[\text{NV}^-] = \frac{k_3 k_4}{\Gamma_0 + k_4I} [\text{NV}^{0}]
\end{equation}
\begin{equation}
[\text{NV}^-] + [\text{NV}^{-,*}] + [\text{NV}^0] + [\text{NV}^{0,*}] = [\text{NV}]
\label{eqNV4}
\end{equation}
where $\text{NV}$ is the total number of NV centers, which is a constant. This result is consistent with existing studies on the charge carrier generation of NV center charge cycling, showing that there is a very weak dependence of the light intensity on the relationship between NV$^-$ and NV$^0$\,\cite{aslam_photo-induced_2013}.

%%%%%%
\subsection{Proposed surface reactions for C-H bond activation}

Once the electron-hole pair is generated, the electron-hole pairs can undergo recombination or be captured by other charge traps before they reach the surface. For these procedures, we simply model them as the annihilation of the electron-hole pairs.
\begin{equation}
\ce{h^+ + e^- } \xrightarrow[]{k_5} \ce{(e^--h^+)_{VB}}
\end{equation}

For the ones that reach the surface, they will participate in the photocatalysis through charge transfer to the surface adsorbates. Now, we consider the dynamics of the chemical reaction when there is a continuous flow of holes to the surface under the electric field resulting from the band bending, as discussed in the main text. According to \cite{chakrapani_charge_2007}, the chemical potential of the surface is the oxygen redox couple of the water adsorbate, leading to a charge transfer from the diamond to the surface. The associated chemical equilibrium at the surface is
\begin{equation}
4(\text{e}^-\text -\text{h}^+)_{\text{VB}} + 4\text{O}_2 + 2\text H_2\text O + 4\text{CO}_2 \ce{<=>} 4(\text{h}^+)_{\text{VB}} + 4 \text{HCO}^-_3
\end{equation}
When there is a flow of holes from the diamond to the surface, the concentration of holes on the right-hand side increases. According to Le Chatelier's principle, the equilibrium will shift in the backward direction. As for the microscopic dynamics for a single hole, it yields
\begin{equation}
\ce{h}^+ + \text{HCO}_3^- \xrightarrow[]{k_6} \ce{\cdot OH + CO_2}
\end{equation}
This is one of the key steps in photocatalysis, where the photogenerated hole produces free radicals and accelerates the reaction. In this context, the reverse reaction of the oxygen redox couple in water, upon hole capture, generates the highly reactive hydroxyl radical, which is an extremely active oxidizing agent in the presence of a $\ce{C-H}$ bond. It can, therefore, efficiently capture a hydrogen atom from the $\ce{C-H}$ bond on the surface:
\begin{equation}
\ce{\cdot OH + C-H} \xrightarrow[]{k_7} \ce{H_2O + C\cdot}
\end{equation}
Once $\ce{C-H}$ is activated, leaving the carbon radical on the surface, it will undergo oxidation in the presence of various oxidants, such as the hydroxyl radical and the oxygen molecule. A very likely subsequent reaction would involve the hole-generated hydroxyl radical immediately recombining with the carbon radical, forming an OH-terminated surface:
\begin{equation}
\ce{C\cdot}+ \ce{\cdot OH} \xrightarrow[]{k_8} \ce{C-OH}
\end{equation}

With these reaction steps, we can build the rate equations that ultimately relate the total surface oxidation rate to photon flux-dependent electron-hole generation. We begin by formulating intermediate agents whose concentration remains constant in the steady state, resulting in a vanishing time evolution. First, the time evolution of hole concentration is given by
\begin{equation}
    \frac{d}{dt}[\text{h}^+] = k_4I[\text{NV}^{0,*}] - k_5[\text{h}^+][\text{e}^-] - k_6[\text{HCO}_3^-][\text{h}^+] = 0
\end{equation}
with $[\text{h}^+] = [\text{e}^-]$, this expression is reduced to
\begin{equation}
    0 = k_4I[\text{NV}^{0,*}] - k_5[\text{h}^+]^2 - k_6[\text{HCO}_3^-][\text{h}^+]
    \label{eqh}
\end{equation}
Next, we examine the population of the hydroxyl radical
\begin{equation}
    \frac{d}{dt}[\cdot \text{OH}] = k_6[\text{HCO}_3^-][\text{h}^+] - k_7[\cdot \text{OH}][\ce{C-H}]_s -k_8[\text{C}\cdot][\cdot \text{OH}] = 0
\end{equation}
with $[\cdot]_s$ denoting the two-dimensional surface concentration and the carbon radical,
\begin{equation}
    \frac{d}{dt}[\cdot\text{C}] = k_7[\cdot \text{OH}][\ce{C-H}]_s - k_8[\text{C}\cdot][\cdot \text{OH}] = 0
\label{eqCd}
\end{equation}
With the hole being transferred to the surface, the surface $\ce{C-H}$ bond and the adsorbate associated with the hydrogen termination will be consumed. Their time derivatives can be expressed as
\begin{equation}
    \frac{d}{dt}[\ce{C-H}]_s = -k_7[\cdot \text{OH}][\ce{C-H}]_s
    \label{eqCH}
\end{equation}
\begin{equation}
    \frac{d}{dt}[\text{HCO}_3^-] = -k_6[\text{HCO}_3^-][\text{h}^+]
    \label{eqHCO}
\end{equation}
Using Eq.\,(\ref{eqCd}) and Eq.\,(\ref{eqHCO}), we can solve the time evolution of the $\ce{C-H}$ bond on the surface
\begin{equation}
    [\ce{C-H}]_s(t) = [\ce{C-H}]_{s,0} \exp(-k_6[\text{h}^+]t) \propto \exp(-Kt)
\end{equation}
Here, it verifies the pseudo-first-order reaction for the $\ce{C-H}$ bond activation mentioned in the main text. The validity of modeling the total surface oxidation as pseudo-first order relies on the assumption that the hole concentration reaching the surface is constant. In our experiment, although the total band bending energy decreases throughout the entire oxidation procedure, this remains a valid approximation since the resulting electric fields during the entire process stay within the same order of magnitude. Therefore, it is reasonable to assume that the hole flow to the surface is nearly constant during photocatalytic oxidation.

To relate the reaction rate $k$ to the photon flux, we now examine how the hole concentration depends on the photon flux as determined by Eq. (\ref{eqh}). In this context, considering the laser power we use (in the mW range) and the pillar geometry, where the wave guiding effect excites the defects along the pillar with comparable efficiency, we assert that the defects responsible for electron-hole pair generation (for example, NV centers) are driven above saturation. According to Eq. (\ref{eqNV1})-(\ref{eqNV4}), $[\text{NV}^{0,*}]$ can be considered independent of photon intensity (flux) and is given by a fraction of the total NV concentration as $[\text{NV}^{0,*}] = \alpha_0[\text{NV}]$, where $\alpha_0$ is a constant that depends only on the wavelength. Therefore, Eq.\,(\ref{eqh}) can be written as
\begin{equation}
    k_5[\text{h}^+]^2 + k_6[\text{HCO}_3^-][\text{h}^+] = k_4\alpha_0[\text{NV}]I
\end{equation}
From this expression, we can obtain two limits for the dependence of the photon flux (intensity). When electron-hole pair recombination and charge trapping are predominant, we have
\begin{equation}
    k_5[\text{h}^+]^2 \gg k_6[\text{HCO}_3^-][\text{h}^+].
\end{equation}
In this regime, the dependence of the steady-state hole concentration on photon flux is approximated as
\begin{equation}
    [\text{h}^+] = \sqrt{\frac{k_4\alpha_0[\text{NV}]}{k_5}I},
\end{equation}
and the total reaction rate $k$ can be expressed as
\begin{equation}
    k = k_6\sqrt{\frac{k_4\alpha_0[\text{NV}]}{k_5}I} \propto I^{0.5} \propto \Phi^{0.5} 
\end{equation}
This corresponds to a high laser power regime, where most of the photon-generated electrons and holes undergo recombination, and the defects are located far from the surface. As a result, only a small fraction of the photo-generated carriers participate in surface photocatalysis. This regime aligns well with our experimental parameters, where the applied power is above saturation, and the NV centers responsible for photocatalysis are the native NV centers, which are typically situated deeper in the pillar compared to the SiV centers that we implant into the diamond. Therefore, the photon flux dependence we reveal in the main text serves as a strong indication of the photocatalytic nature of the dynamics. 

The other limit of photon dependence is low laser power or defects close to the surface regime where 
\begin{equation}
    k_5[\text{h}^+]^2 \ll k_6[\text{HCO}_3^-][\text{h}^+],
\end{equation}
and the charge transfer to the surface is predominant. The photon flux dependence of the steady state hole concentration becomes
\begin{equation}
    [\text{h}^+] = \frac{k_4\alpha_0 [\text{NV}]}{k_6[\text{HCO}_3^-]}I
\end{equation}
With the high power limit, our model reveals the characteristic photon flux dependence of the photocatalytic reaction rate\,\cite{buglioni_technological_2022}. 
Note that for even lower laser power, where the charge source defects (NV centers) are driven below saturation, the population of the excited state of NV$^0$ will become photon-intensity dependent and, at this regime, can be approximated as $[\text{NV}^{0,*}] = \alpha_0'[\text{NV}]I$. This makes the total photon flux dependence quadratic.

From these expressions, we can see that the total reaction rate also depends on the local NV center concentration. This explains the deviation in the fitted reaction rate we observed from different pillars under the same laser power and wavelength. It also justifies the need to average data from multiple pillars to reveal a meaningful dependence of the reaction rate on laser power and wavelength. This is because the local fluctuations in the defect distribution should be averaged across the diamond.

%%%%%%
\subsection{Order of magnitude estimate of the macroscopic rate constant}

To further verify the validity of our model, we estimate the order of magnitude for the macroscopic reaction rate constant in the observed scaling order regime of the photon flux dependence, i.e. $k\propto \Phi^{0.5}$. This regime corresponds to a photocatalysis procedure under high laser power, where the defect is driven above saturation and only a small fraction of photogenerated holes are transferred to the surface because of the high excitation power and the deep location of these defects.

We start by estimating the charge carrier generation rate of an NV center driven above saturation. According to \cite{siyushev2019photoelectrical}, the generation rate $\Gamma_{ion} \sim 10^6$\,Hz. To estimate the number of NV centers that can contribute to the procedure, we first take the concentration of native NV centers in the diamond as provided by the manufacturer, $[\text{NV}] < 0.03\,\text{ppb} = 5.4\times10^{18}\,\text{m}^{-3}$. If we consider that only the NV centers in the truncated cone part of the tip can efficiently contribute to the procedure, this number $N_{\rm eff}$ can be calculated by
\begin{equation}
    N_{\rm eff} = [\text{NV}]\cdot\frac{\pi h}{3}\left(r_1^2 + r_2^2 + r_1r_2\right)
\end{equation}
Using $r_1 =$ \qty{0.35}{\um}, $r_2 = $\qty{0.5}{\um} and $h = $\qty{1}{\um} which are typical for our nanopillar structure, we obtain $N_{\rm eff} < 3.1$. The total macroscopic reaction rate can be estimated by the following expression:
\begin{equation}
    k \sim \frac{\eta N_{\rm eff}\Gamma_{ion}}{N_{\rm C}}
\end{equation}
where $\eta$ is the fraction of photogenerated holes involved in photocatalysis, and $N_{\rm C}$ is the total number of carbon sites on the top surface of the pillar, with $N_{\rm C} = [\ce{C}]_{s}\cdot\pi r_1^2 \approx 6.16\times10^{6}$ (using $[\ce{C}]_{s} = 16\,\rm{nm}^{-2}$ for the diamond surface (001). To match the observed order of magnitude of the reaction rate $k\sim10^{-3}$\,Hz, it imposes $\eta\sim1\%$, which corresponds to the regime we expect, where a small fraction of the photogenerated holes participates in photocatalysis.

%%%%%%%%%%%%
\section{Derivation of the time dependent adiabatic charge state population for modeling of the PL trace}
\label{SI_Der}
In this section, we derive the expression for the thermodynamically stabilized charge state population to which SiV$^-$ PL is proportional, starting from the formalism of formation energy\,\cite{defo_charge-state_2023}. In its most general form, the population of a given charge state $q$ of a defect $X$, when the Fermi level is close to the adiabatic charge state transition level between the $q$ and $q+1$ states, is defined as
\begin{equation}
    P_{X^q} = \frac{\exp\left[ -\beta\Delta H_X(q,E_F)\right]}{\exp\left[ -\beta\Delta H_X(q,E_F)\right] + \exp\left[ -\beta\Delta H_X(q+1,E_F)\right]}
\end{equation}
where $\Delta H_X(q,E_F)$ is the formation energy of the charge state $X^q$ at a given Fermi level $E_F$. Using the following relations that connect the formation energy to the charge state transition level $E_{q/q+1}$:
\begin{equation}
  \Delta H_X(q,E_F) = \Delta H_X(q,0) + qE_F  
\end{equation}
\begin{equation}
  E_{q/q+1} = \Delta H(q,0) - \Delta H(q+1,0)
\end{equation}
We obtain immediately for a single color center
\begin{equation}
    P_{X^q} = \frac{1}{1+\exp\left[\beta(E_{q/q+1} - E_F) \right]}.
\end{equation}
When the collected PL $I_{X^q}$ is only from the emission of a single $X^q$,  we can determine directly $I_{X^q}\propto P_{X^q}$.

However, when the probed pillar contains several emitters at different depths, the total collected PL intensity from charge state $X^q$ is given by a linear combination of the charge state populations of individual emitters:
\begin{equation}
    I_{X^q} = \sum_{i = 1}^{N} I_iP_{X^q}^{i} = \sum_{i = 1}^{N} \frac{I_i}{1+\exp\left[ \beta(E_{q/q+1}^{i} - E_F)\right]}
    \label{eqIexc}
\end{equation}
where $I_i$ is the $X^q$ PL contribution from the $i^{\rm th}$ emitter in the nanopillar when it is fully in charge state $q$ and $E_{q/q+1}^{i}$ is the charge state transition level at the depth of $i^{\rm th}$ emitter.
When the spread in $E_{q/q+1}^{i}$ is small compared to the thermal broadening of Fermi level ($\sim~k_BT$), we expand Eq.\,(\ref{eqIexc}) around the ensemble average charge state transition level $\overline E_{q/q+1}$.
Expressing $E_{q/q+1}^{i} = \overline{E}_{q/q+1}+\delta_i$ with $\overline \delta_i = 0$:
\begin{equation}
    I_{X^q} = \sum_{i = 1}^{N} \frac{I_i}{1+\exp\left[ \beta(\overline E_{q/q+1} + \delta_i - E_F)\right]} = \frac{1}{1+\exp\left[ \beta(\overline E_{q/q+1} - E_F)\right]} \sum_{i = 1}^{N} I_i\left[1 - \left( \frac{\beta\delta_i}{\exp\left[ \beta( E_F - \overline E_{q/q+1} )\right] + 1}\right)\right].
\end{equation}
Under the condition of small energy spread, where $\beta|\delta_i| \ll 1$, the higher-order terms become negligible. The total PL intensity is then approximated as:
\begin{equation}
    I_{X^q} \approx \frac{1}{1+\exp\left[ \beta\left(\overline E_{q/q+1} - E_F\right)\right]} \sum_{i = 1}^{N} I_i,
\end{equation}
which simplifies to a single Fermi-Dirac distribution scaled by the constant total intensity $\sum_{i = 1}^{N} I_i$. 

The condition $\beta|\delta_i| \ll 1$ holds for our 80\,keV-$^{28}$Si-implanted sample. 
SRIM simulations show a depth spread of $\Delta d \sim 10\,$nm for the resulting SiV centers, inducing an energy spread of $\delta_i\sim \pm5$\,meV at a mean depth of 50\,nm (according to the band bending calculation detailed in section-\ref{SI_BB}).
This spread is significantly smaller than the thermal energy $k_BT$ at room temperature ($\sim 25$\,meV).
Consequently, for nanopillars containing a micro-ensemble of SiV centers with minimal depth variations, the total measured SiV$^-$ PL intensity follows:
\begin{equation}
    I_{\text{SiV}^-} \propto P_{\text{SiV}^-} = \frac{1}{1+\exp\left[\beta(E_{-/0} - E_F)\right]}
\end{equation}
where $E_{-/0}$ now signifies the charge state transition level at the averaged position of the SiV ensemble in the nanopillar, consistent with our main-text assignment. 

As defined in the main text, the difference between the charge state conversion level and the Fermi level can be written as
\begin{equation}
E_{q/q+1} - E_F = \mu_e - E_g -E_{ea} + \Delta_{q/q+1} - \delta_{BB},
\label{diffE}
\end{equation}
with $\mu_e$ the surface chemical potential, $E_g$ the diamond band gap, $E_{ea}$ electron affinity, $\Delta_{q/q+1}$ the difference between the charge state transition level $E_{q/q+1}$ and the valence band maximum, and $\delta_{BB}$ the band bending energy from the surface to the average depth of the color center.

As discussed in the main text, the time-dependent term during the laser illumination is the electron affinity (EA) $E_{ea}$, which can be expressed as a weighted average between the initial and final EA, $E_{ea}^{\rm H}$ and $E_{ea}^{f}$. 
Combining this with the pseudo-first order decay of the $\ce{C-H}$ bond concentration, we can explicitly write the time-dependent EA for any arbitrary initial starting time $t_0$ of the PL measurement as
\begin{equation}
    E_{ea}(t) = E_{ea}^{\rm H}e^{-k(t+t_0)} + E_{ea}^{f}\left(1 -e^{-k(t+t_0)}\right)
\label{eqEt}
\end{equation}
By substituting the $E_{ea}$ in Eq. (\ref{diffE}) with Eq. (\ref{eqEt}), we can express the total time-dependent charge state population in the form of
\begin{equation}
    P_{X^q}(t) = \frac{1}{1+\exp\left[A + B\exp(-Kt)\right]}
\end{equation}
with 
\begin{equation}
    A = \beta\left( \mu_e-E_g+\Delta_{q/q+1}-E_{ea}^f - \delta_{BB}\right),
\end{equation}
\begin{equation}
    B = \beta\left(E_{ea}^f-E^H_{ea}\right)e^{-Kt_0}
\end{equation}
and their sum
\begin{equation}
    A+B = \beta(\mu_e - E_g  +\Delta_{q/q+1}-E_{ea}(t_0)-\delta_{BB}) = \beta(E_{q/q+1}^{0} - E_F),
\end{equation}
which signifies the initial difference between the position of the charge state transition level and the Fermi level, as stated in the main text.

%%%%%%%%%%%%
\section{NV PL spectral signature observed in SiV samples}
\label{SI_NVPL}
Here, we present the PL spectra obtained from our SiV samples to illustrate the presence of native NV through its typical spectroscopic signature. 

 \begin{figure}[h]
    \centering
    \includegraphics[width=115mm]{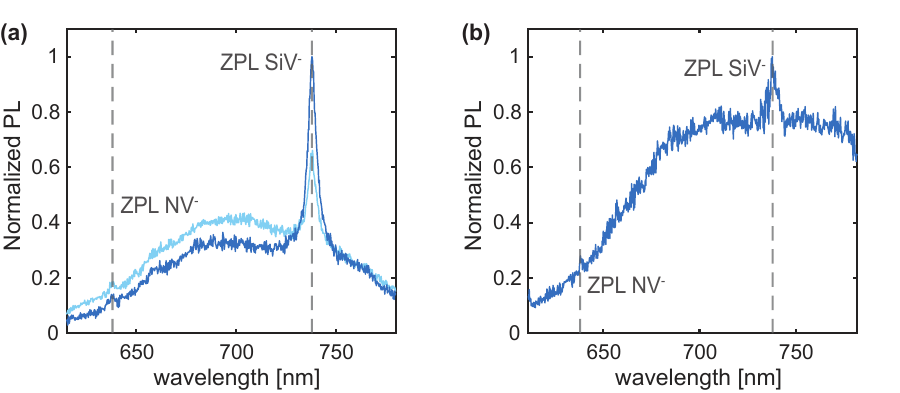}
    \caption{(a) PL emission spectra acquired on an SiV bulk sample before the nanopillar is fabricated. Two spectra are acquired in regions with different SiV implantation doses. Dark blue curve represents the spectrum acquired in the region with implantation parameter the same as the sample we preformed the experiment in the main text. The light blue curve shows the spectrum acquired in the region with three times less concentrated implantation dose. 
    Both regions manifest a typical NV PL spectra background with comparable intensity, benchmarking the intensity from native NV density in such diamond.
    (b) One exemplary background PL spectrum taken in one pillar of the SiV sample we used for all the experiment presented in the main text, also showing a typical NV spectral signature. The confocal point is set to be off the SiV PL intensity maximum.}
    \label{figS6}
\end{figure}

%%%%%%%%%%%%
\section{Conversion observed on NV sample}
\label{SI_NVconv}

Since the optical charge state conversion discussed in this work pertains to reactions on the diamond surface, it is not limited to SiV centers and should also be observable with other defects in diamond.
Therefore, we performed optical charge state conversion on a diamond nanopillar sample implanted with single shallow ($\sim15\,$nm) NV centers.
However, probing the charge state conversion of NV centers in diamond by measuring the PL is not as straightforward as it is for the SiV centers, since the PL of the neutral and negative charge states (NV$^0$ and NV$^-$) has a broad region of overlap. 

\begin{figure}[h]
    \centering
    \includegraphics[width=168mm]{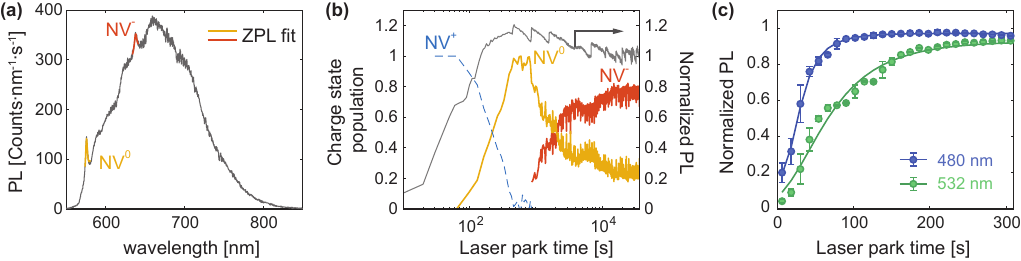}
    \caption{(a) Total NV photoluminescence (PL) collected from 575\,nm to 800\,nm as a function of the blue (480\,nm) and green (532\,nm) laser park time on the same diamond nanopillar with single NV center in two separate H-termination runs, both at a power of 0.7\,mW. Each PL time trace is fitted by the charge state conversion model in solid lines.
    (b) The ZPL intensity inferred charge state population time evolution of NV$^+$ (blue dashed line), NV$^0$ (yellow solid line) and NV$^-$ (red solid line) plotted together with the total time dependent PL trace (gray solid line). The total PL is normalized to its intensity at long term limit.
    (c) One exemplary PL spectrum during the optical charge state conversion of NV center under 532\,nm excitation at 0.7\,mW. The background subtracted ZPL of NV$^0$ and NV$^-$ are fitted by Lorentzian functions. 
    }
    \label{figS7}
\end{figure}

In order to disentangle the PL contributions from NV$^0$ and NV$^-$, we performed PL spectroscopy analysis during the laser illumination.
The emitted PL is directed to a beamsplitter, which sends 50$\%$ of the PL to a spectrometer for spectroscopic analysis and the remainder to an APD. 
Each spectrum is recorded every 30 seconds with an integration time of 30 seconds.
One exemplary spectrum is shown in Fig. \ref{figS7} (b). where both NV$^0$ and NV$^-$ zero-phonon lines (ZPL) are clearly visible.
The PL intensity inferred charge state population of NV$^0$ and NV$^-$ is estimated by the integrated ZPL intensity from Lorentzian fitting of the corresponding ZPL after subtracting the Phonon Sideband (PSB) background, as shown in Fig. \ref{figS7} (a).
Since the spread of the Fermi level at room temperature is much smaller than the difference between the two charge state transition levels ($E_{0/+} - E_{-/0} \gg k_BT$), we consider that there are at most two charge states coexisting at each stage.
In an early stage, the low NV$^0$ PL signal is attributed to a dominant population of NV$^+$ charge states. The subsequent increase in NV$^0$ PL is due to charge conversion from NV$^+$ to NV$^0$. Once both ZPLs of NV$^0$ and NV$^-$ are visible, we set the NV$^+$ population to zero and extract the respective ZPL intensities from the Lorentzian fitting, then normalize their sum at each time moment to obtain the relative contribution of each charge state.
An example of the resulting charge-state population evolutions is shown in Fig. \ref{figS7} (b), and the saturated population ratio between NV$^0$ and NV$^-$ corresponds to the equilibrium charge-state distribution of NV under continuous laser excitation of 532\,nm due to the NV charge cycling\,\cite{aslam_photo-induced_2013}.

In Fig. \ref{figS7} (c), we show the photoluminescence (PL) over the laser park time from a single nitrogen vacancy (NV) center located in a diamond nanopillar structure. 
The same NV center was studied in two separate runs, each following a fresh hydrogen termination. In the first run, the NV was illuminated with a 480\,nm laser and in the second run with a 532\,nm, both at a power of 700\,$\mu$W.
By fitting the initial part of the PL time evolution when PL increase is mainly caused by the charge state transition from NV$^+$ to NV$^0$ (according to the result from Fig. \ref{figS7} (b)), with the same model presented in the main text, we extracted the reaction rate constant $k=15.6\pm10.8\,$mHz for green laser illumination and $k = 30.9\pm11.4\,$mHz for blue illumination.
We can see that the $k$ extracted from a sample with a near-surface NV center is larger than the one extracted from the SiV sample by an order of magnitude, for even much lower laser power.
It marks the significant role of NV centers in this photocatalytic surface oxidation procedure.
According to our proposed model, it is as expected: When there is an NV center close to the surface, photogenerated holes are much more likely to be driven to the surface, yielding a larger fraction of photoholes participating in photocatalysis reactions, which efficiently enhances the reaction rate. 
Moreover, further enhancement of the reaction rate constant $k$ for illumination at 480\,nm indicates that defects other than the NV center contribute to photohole generation upon excitation at this photon energy. This could be additional support for the divacancy center involvement in photocatalysis as mentioned in the main text.

%%%%%%
\section{Calculation of band bending in the diamond nanopillar}
\label{SI_BB}
In this section, we show the details of the band bending calculation for our diamond nanopillar, which we use to estimate the depth range in which native diamond defects can contribute to surface photocatalytic oxidation.
We begin by identifying the dominant dopants. Our electronic grade diamond contains nitrogen at a concentration of 5\,ppb (equivalent to $N_d = 0.9\times10^{21}\text{m}^{-3}$). 
The concentration of boron is below <1\,ppb and is therefore neglected.
Here, we apply the classical depletion approximation, which is considered a good approximation because:
(i) the dopant density is uniform and very low such that the depletion depth greatly exceeds the Debye length, and (ii) while the mobile charge carriers are negligible, all donors are ionized at room temperature inside the charged layer. Therefore, the space-charge density $\rho_c$ can be well described solely by fixed ionized donors, where $\rho_c = +eN_d$.

\begin{figure}[h]
    \centering
    \includegraphics[width=140mm]{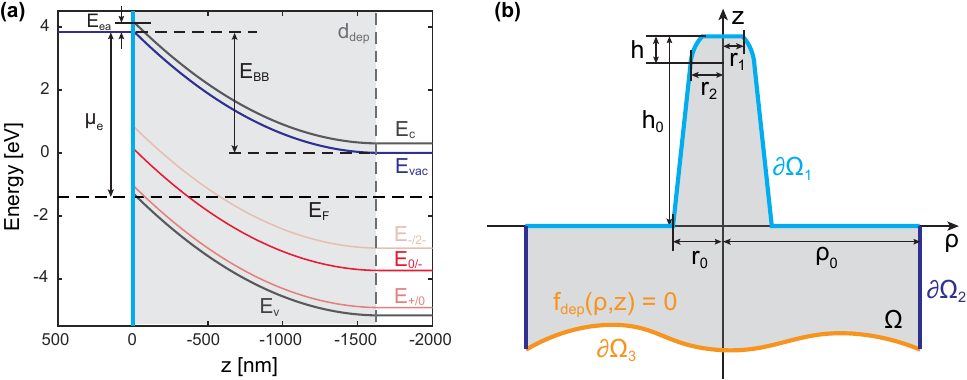}
    \caption{(a) An exemplary picture for a upward band bending for a flat diamond surface under the depletion approximation with a negative electron affinity with respect to the vacuum level $E_{\rm vac}$ (blue solid line), corresponding to a typical case of a H-terminated surface. The upward band bending energy $E_{\rm BB}$ is marked for the vacuum level $E_{\rm vac}$. The conduction band minimum $E_c$ and valence band maximum $E_{\rm v}$ of diamond are depicted with gray solid lines.
    The depletion layer from the surface to the depletion depth $d_{\rm dep}$ is marked as gray shaded area. Different SiV charge state transition levels $E_{2-/-}$, $E_{-/0}$ and $E_{+/0}$ are shown according to the calculated value in \cite{gali_ab_2013}.
    (b) The relevant domaine $\Omega$ that we used to numerically solve the Poisson equation in our pillar structure in cylindrical coordinate, defined by three sections of boundaries: $\partial\Omega_1$,$\partial\Omega_2$ and $\partial\Omega_3$.}
    \label{FigS8}
\end{figure}

Using these conditions, we can first solve the Poisson equation for a flat surface of a bulk diamond. The sketch of the problem is illustrated in Fig. \ref{FigS8} (a).
Since the band bending is caused by a different chemical potential at the surface from the bulk Fermi level, according to \ref{FigS8} (a), the amount of band bending shift $E_{BB}$ can be written as 
\begin{equation}
    E_{\rm BB} = \mu_e - E_{ea} - (E_c - E_F)\big|_{\infty}
\end{equation}
with $\mu_e$ the chemical potential of the surface adsorbate, $E_{ea}$ the electron affinity, and $(E_c - E_F)|_{\infty}$ the bulk limit of the difference between the conduction band minimum $E_c$ and the Fermi level $E_F$.
Taking the vacuum level as an example, the problem to be solved is essentially a one-dimensional Poisson equation in an undefined domain $\Omega \in [-d_{\rm dep},0]$:
\begin{equation}
\nabla^2 E_{\rm vac} = \frac{d^2}{dz^2}E_{\rm vac} =  - \frac{\rho_c}{\varepsilon_0 \varepsilon_r} = -\frac{e N_d}{\varepsilon_0 \varepsilon_r}
\end{equation}
where $\varepsilon_r = 5.7$ is the diamond relative permittivity, and $\varepsilon_0$ is the vacuum permittivity, with a Neumann boundary condition at the depletion depth $d_{\rm dep}$
\begin{equation}   
\frac{dE_{\rm vac}}{dz}\bigg|_{-d_{\rm dep}} = 0,
\end{equation}
then we impose two Dirichlet boundary conditions by setting the bulk limit of $E_{\rm vac}$ to be zero
\begin{equation}
    E_{\rm vac}(0) = E_{BB},
\end{equation}
\begin{equation}
    E_{\rm vac}(-d_{\rm dep}) = 0,
\end{equation}
we can solve $E_{\rm vac}(z)$ and $d_{dep}$ at the same time. The analytical expression of $E_{\rm vac}(z)$ can be written as 
\begin{equation}
E_{\rm vac}(z) = \frac{E_{BB}}{d_{\rm dep}^2}\left( z + d_{\rm dep} \right)^2
\end{equation}
with
\begin{equation}
    d_{\rm dep} = \sqrt{\frac{2\epsilon_0\epsilon_rE_{BB}}{eN_d}}
\end{equation}
With this expression, we obtain the depth of depletion for the H-terminated surface for a flat surface as $d_{\rm dep}$\qty{1.7}{\um} and the O-terminated surface as $d_{\rm dep}\sim$\qty{1.2}{\um} (using $E_{ea}^{O} = 1.7\,$eV\,\cite{maier2001electron}). For both terminations, depletion depth is greater than the largest size of our pillar structure. This means that during the entire surface oxidation procedure, the whole pillar is depleted, and the band bending extends up to the bulk part of the diamond.

Now, we focus on the band-bending calculation in our pillar structure. Here, we again choose the vacuum level to solve the spatial evolution of the energy bands. We again set the bulk limit of $E_{\rm vac}$ at zero. Since the nanopillar structure shown in Fig. \ref{figS1} has a cylindrical symmetry, we chose to solve the Poisson equation in the cylindrical coordinate. Applying the symmetry $\partial_\theta E_{\rm vac} = 0$, the question to be solved becomes
\begin{equation}
\nabla^2 E_{\rm vac} = \frac{1}{\rho}\frac{\partial}{\partial\rho}\left( \rho\frac{\partial E_{\rm vac}}{\partial\rho}\right) + \frac{\partial^2 E_{\rm vac}}{\partial\rho^2} = -\frac{e N_d}{\varepsilon_0 \varepsilon_r}, \text{ in domain }\Omega.
\end{equation}
The domain $\Omega$ is shown in Fig. \ref{FigS8}. The boundary $\partial\Omega$ that defines this domain is divided into three different parts with different applied boundary conditions: (i) the surface of the diamond denoted as $\partial\Omega_1$ with a Dirichlet boundary condition given by the surface Fermi level pinning:
\begin{equation}
    E_{\rm vac}|_{\partial\Omega_1} = E_{\rm BB};
\end{equation}
(ii) the bulk radial limit denoted as $\partial\Omega_2$ which is chosen to be in a radial position far from the base of the pillar $\rho_0\gg r_0$ and the band bending in diamond is considered to be the same as a flat surface. This yields a Dirichlet boundary condition as
\begin{equation}
    E_{\rm vac}(\pm\rho_0,z) = \frac{E_{BB}}{d_{\rm dep}^2}\left( z + d_{\rm dep} \right)^2;
\end{equation}
(iii) The bulk depth limit signifies the depletion depth denoted as $\partial\Omega_3$, described by a curve $f_{dep} (\rho,z) = 0$, which is undefined. It needs to be solved at the same time as the total band bending by applying the following Dirichlet and Neumann boundary conditions:
\begin{equation}
    E_{\rm vac}|_{\partial\Omega_3} = 0
\end{equation}
\begin{equation}
    \frac{\partial E_{\rm vac}}{\partial n}\bigg|_{\partial\Omega_3} = 0
\end{equation}
where $\frac{\partial}{\partial n}$ denotes the derivative along the normal direction of the boundary. 
The problem is then solved numerically using the commercially available software Mathematica. The dimensions of the nanopillar are chosen according to its SEM image shown in Fig. \ref{figS1}. The pillar apex has a truncated parabolic shape with a top radius $r_1 \sim$\qty{350}{\nm}, a bottom radius $r_2 \sim $\qty{500}{\nm}, and a height $h\approx$\qty{700}{\nm}. The base of the pillar is an isosceles trapezoid, \qty{3.4}{\um} in height, with an angle of $\sim83^\circ$ at the bottom. The bulk radial limit $\rho_0$ is chosen as half of the average distance between the pillars, which is $\sim$\qty{5}{\um}.

The calculated results of the vacuum level $E_\text{v}$ bending for the RT H-terminated diamond are presented in Fig.\ref{fig3}\,(d), with color coding used to represent the potential profile. The white arrows indicate the resulting electric field lines, where the length of the arrows corresponds to the field strength. Our numerical calculation shows that, due to the geometry of the pillar, an upward band bending is observed throughout its volume, creating an effective potential that facilitates hole trapping and propagation toward the surface. Although the toy model is oversimplified and may have quantitative inaccuracies, it qualitatively captures the key feature of interest, supporting our theoretical hypotheses.

%%%%%%%%%%%%
\section{Additional observation of PL dynamics at low temperature (LT)}
\label{SI_LT}

 \begin{figure}[h]
    \centering
    \includegraphics[width=150mm]{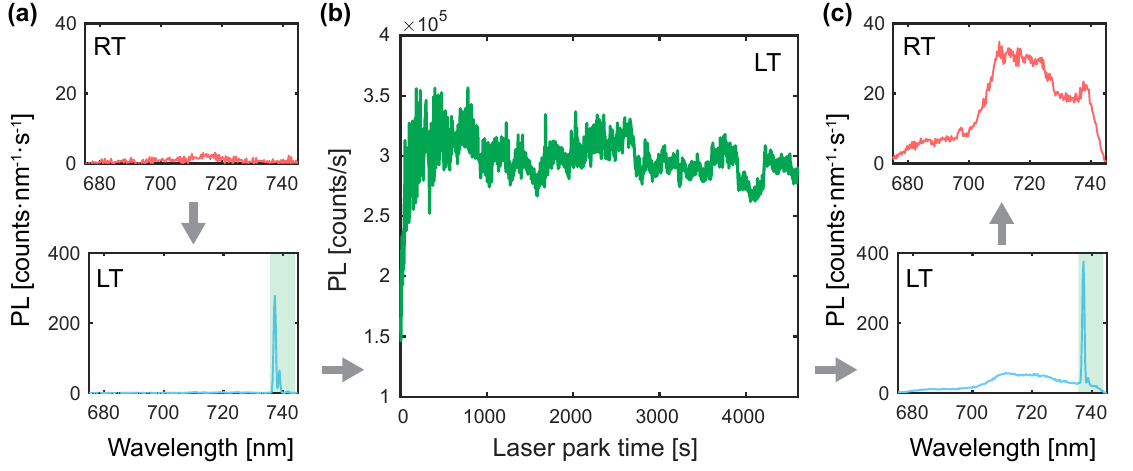}
    \caption{(a) The PL spectra of nanopillar containing SiVs a H-terminated surface at room temperature (RT, upper panel, ambient conditions) and at low temperature (LT, lower panel, 7\,K in a dry cryostat) before prolonged visible laser illumination. 
    (b) The PL trace collected around the SiV ZPL wavelength range (733\,nm to 747\,nm, highlighted in green in the lower panel of (a) and (c)), recorded under continuous wave (CW) 515\,nm laser illumination at 14\,mW at LT. 
    (c) PL spectra at LT (Lower panel) and RT (upper panel) acquired from the same nanopillar after 1.5\,hours of CW 515\,nm laser illumination at 14\,mW under LT conditions.}
    \label{figS33}
\end{figure}
For a H-terminated nanopillar containing SiVs, we have mentioned in the main text that the SiV$^-$ charge state can be partially recovered under cryogenic conditions (typically 7\,K in our case), as shown in Fig.\,\ref{figS33}\,(a). 
Here, we present additional observations of the behavior of the H-terminated surface under laser illumination at cryogenic environment.
After cooling the sample to cryogenic conditions, we illuminated the same pillar with 515\,nm continuous-wave laser illumination at 14\,mW. for 1.5 hours. 
The resulting PL intensity trace within SiV$^-$ ZPL region is shown in Fig.\,\ref{figS33}\,(b). 
A rapid initial increase in PL is observed, followed by a plateau where the intensity remains stable for the rest of the illumination period.

To investigate the origin of this PL change, we take another PL spectrum (see Fig.\,\ref{figS33}\,(c)) after 1.5 hours illumination. 
The comparison reveals that the increased PL is not due to enhanced SiV$^-$ emission, as no significant change is observed in the ZPL region.
Instead, a noticeable broad-band spectral feature appears after the prolonged exposure of the green laser. This spectral feature persists upon warming the sample back to room conditions, and the SiV$^-$ ZPL remains absent, indicating that the SiV charge state has not been restored. 
However, according to Fig.\,\ref{fig4} in the main text, further laser illumination at room temperature does eventually lead to SiV$^-$ charge state recovery.

These observations suggest that the optical charge state conversion via laser-induced surface oxidation of the H-termination does not proceed under cryogenic conditions.
Interestingly, our data reveal the emergence of a broad emission feature under laser illumination on the H-terminated surface at low temperature, which may indicate the deposition of some form of contaminant. This phenomenon has not been observed for the O-terminated surface in our previous studies.
The exact origin of this contaminant remains unknown, but these results highlight the need for caution when operating H-terminated diamond in a cryogenic environment. The impact on the shallow color centers' properties as well as possible mitigation approaches is still worth further investigation.

%%%%%%%%%%%%
\section{Fluorescence measurement for determine charge state population in Fig.\,\ref{fig4}}
\label{SI_fig4}

\begin{figure}[h]
    \centering
    \includegraphics[width=150mm]{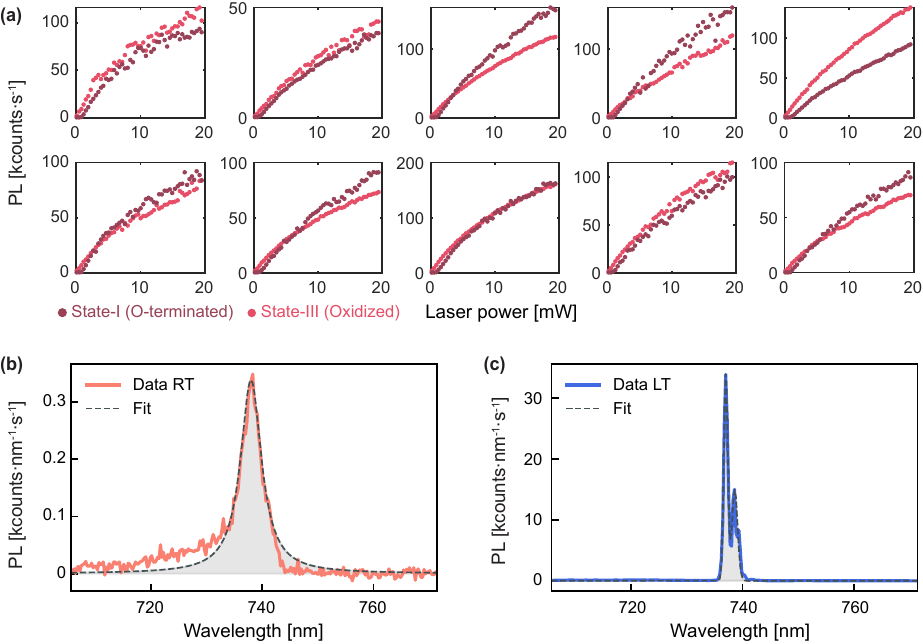}
    \caption{(a) Saturation curves of the SiV$^-$ ZPL under 515\,nm excitation for ten nanopillars, measured under O-terminated (State-I, dark red) and laser-oxidized (State-III, light red) surface at room temperature. The linear-like dependence up to 20\,mW and nearly identical low-power PL indicate comparable SiV$^-$ charge state populations at state-I and -III as expected.
    (b) Representative ZPL spectrum at room temperature fitted with a Lorentzian profile. The gray area indicates the total ZPL emission intensity extracted from the fitting.
    (c) Representative ZPL spectrum at low temperature (7\,K), fitted with a multi-Gaussian profile to account for spectrometer resolution limitations, with the gray area as the total ZPL emission intensity.}
    \label{figS9}
\end{figure}

In this section, we describe the method used to extract SiV ZPL intensity $I_{\text{SiV}^-}$, which is used to determine the relative charge state population shown in Fig.,\ref{fig4} of the main text.
Since each nanopillar in the sample contains several SiV centers at varying depths, the average charge state population of a given pillar at different surface conditions can be inferred from the ZPL intensity under fixed excitation conditions. 
This approach is justified by the near-linear behavior of the saturation curve for a micro-ensemble of SiVs.
We verify this by taking the saturation curve of the SiV ZPL (PL collected from 733 to 747\,nm) under CW 515\,nm off-resonant excitation for ten nanopillars, both with an O-terminated surface and a laser oxidized H-terminated surface, all at room temperature (RT). As shown in Fig.\,\ref{figS9}\,(a), the PL does not saturate up to 20\,mW and the low-power PL intensities are nearly identical for both surface terminations, which is consistent with the expected similar SiV$^-$ populations in these two states.

Based on this, we determine the relative SiV$^-$ population for any given surface termination (at either LT or RT) by taking the ratio of the integrated ZPL intensity of that state to that of State-I (O-terminated) (at LT or RT respectively), under identical excitation conditions.
The integrated ZPL intensity is extracted by fitting the ZPL spectra recorded under CW 515\,nm excitation at 8\,mW with a 30\,s integration time. Representative spectra and their corresponding fits at RT and LT are shown in Fig.\,\ref{figS9}\,(c) and (d). 
To account for minor variations in collection or excitation efficiency at different measurement stages, we normalize all intensities to the ZPL signal collected from a reference single SiV sample, mounted adjacent to the investigated sample.
Specifically, for RT spectra, we use a Lorentzian function:
\begin{equation}
f(x)=\frac{I_0 \gamma^2}{\gamma^2 +(x-x_0)^2},
\end{equation}
where $I_0$ denotes the peak amplitude, $x_0$ represents the averaged center wavelength, and $\gamma$ is the full width at half maximum (FWHM).
For LT spectra,  we employ a Gaussian function:
\begin{equation}
  f(x) = I_0\exp\left(-\frac{(x-x_0)^2}{2 \sigma^2}\right) 
\end{equation}
with $\sigma$ representing the linewidth. The reason for using the Gaussian function for LT spectra is that the true linewidth of SiV at 7\,K is smaller than the resolution of the spectrometer; therefore, the measured spectra can be approximated by a Gaussian function.

\bibliographystyle{apsrev4-2}
\bibliography{SI.bib}